\documentclass[epj]{svjour}
\usepackage{graphics}
\usepackage{graphicx}
\usepackage{amsmath}
\usepackage{amsfonts}
\usepackage{mathtools,cuted}
\usepackage{lipsum}

\newcommand{\beq}{\begin{equation}}
\newcommand{\eeq}{\end{equation}}
\newcommand{\beqa}{\begin{eqnarray}}
\newcommand{\eeqa}{\end{eqnarray}}
\def\lsim{\raise0.3ex\hbox{$<$\kern-0.75em\raise-1.1ex\hbox{$\sim$}}}
\def\gsim{\raise0.3ex\hbox{$>$\kern-0.75em\raise-1.1ex\hbox{$\sim$}}}

\begin{document}
\title{In-medium hadronization of heavy quarks and its effect on charmed meson and baryon distributions in heavy-ion collisions}
\author{Andrea Beraudo\inst{1}, Arturo De Pace\inst{1}, Marco Monteno\inst{1}, Marzia Nardi\inst{1} and Francesco Prino\inst{1}}
\institute{INFN, Sezione di Torino, via Pietro Giuria 1, I-10125 Torino}
\abstract{We present a new model for the description of heavy-quark hadronization in relativistic heavy-ion collisions in the presence of a reservoir of lighter thermal particles with which recombination can occur leading to the formation of color-singlet clusters. Color neutralization is assumed to occur locally, within the same fluid cell occupied by the heavy quark and it proceeds via the recombination of the latter with light antiquarks (quarks) or diquarks (anti-diquarks), which are assumed to be present in the medium with thermal abundance as effective degrees of freedom around the QCD crossover temperature $T_c$. Typically the resulting color-singlet clusters have quite low invariant mass, in most of the cases below 4 GeV, and in this case they are assumed to undergo an isotropic 2-body decay in their local rest-frame. Heavier clusters are instead fragmented as Lund strings. The possibility of recombination with light diquarks enhances the yields of charmed baryons, in qualitative agreement with recent measurements. The assumption of local color neutralization leads to a strong space-momentum correlation, which provides a substantial enhancement of the collective flow of the final-state charmed hadrons, affecting both their momentum and their angular distributions.}

\authorrunning{A. Beraudo, A. De Pace, M. Monteno, M. Nardi and F. Prino}
\titlerunning{In-medium hadronization of heavy quarks}

\maketitle
\section{Introduction}
Modelling hadron production in any kind of high-energy collisions, from $e^+e^-$ to nucleus-nucleus, requires as an intermediate step some mechanism of color neutralization, grouping the partons arising from the scattering (or multiple scatterings) into color-singlet ``clusters'', which will be the parents of (or in some implementations coincide with) the final hadrons.
In more elementary collisions, like $e^+e^-$ or to a certain extent proton-proton, and in the large-$N_c$ approximation (hence neglecting interference effects, suppressed by $1/N_c^2$ factors, with $N_c\!=\!3$ being the actual number of colors in QCD) it is possible to follow in an unambiguous way the color flow in the hard scattering and in the associated initial (for hadronic beams) and final-state parton-shower evolution. Hence, one can group all the partons (and possibly beam-remnants) in the final state of the shower evolution into color-singlet excited systems denoted as strings~\cite{Andersson:1983ia} and clusters~\cite{Webber:1983if} in the PYTHIA and HERWIG event generators, respectively. The two differ essentially in their invariant mass, clusters being usually lighter than strings, since in HERWIG all gluons at the end of the shower evolution are forced to split into quark-antiquark pairs, breaking the color-connections and giving-rise to the production of many smaller color-neutral systems. Finally, strings and clusters decay into hadrons, in both cases via the excitation of quark-antiquark or diquark-antidiquark pairs from the vacuum. Hence, within such a setup \emph{most valence quarks of the final hadrons} do not arise from the hard event or from the associated parton-shower stage, but are actually \emph{produced} via quantum-mechanical tunnelling \emph{at hadronization}. Furthermore, this approach provides the richest possible information on the event, not only on the production of the hardest particle, but also on the associated soft multiplicity.
On the other hand, if in the collisions of two hadrons one is simply interested in the inclusive production of some hard particle $h$ (for which $p_T$ and/or $M$ are $\gg\Lambda_{\rm QCD}$) one can evaluate the corresponding cross section exploiting the QCD factorization theorem. Considering for instance the production of some charm hadron $h_c$ the result is given by the following convolution
\beq
d\sigma_h=\sum_{a,b,X}f_a(x_1)\,f_b(x_2)\,\otimes d\hat\sigma_{ab\to c\bar cX}\,\otimes D_{c\to h_c}(z)\,,
\eeq
where $f_{a/b}$ are the parton distribution functions of the incoming hadrons, $d\hat\sigma$ is the partonic cross-section and the daughter hadron $h_c$ carries a fraction $z\le 1$ of the light-cone momentum of the parent charm quark. The above fraction is distributed according to the \emph{fragmentation function} $D_{c\to h_c}(z)$, whose integral
\beq
\int_0^1 dz D_{c\to h_c}(z)=f_{h_c}
\eeq
provides the \emph{fragmentation fraction} $f_{h_c}$, i.e. the probability for a charm quark to fragment into a given charmed hadron species. Within the above factorized approach both the fragmentation functions and the fragmentation fractions are assumed to be \emph{universal}, i.e. they can be measured in a given process like $e^+e^-$ annihilation and used in other collision systems, like for instance proton-proton. Furthermore, within this scheme the final heavy-flavour hadron always carries a smaller energy than the parent heavy quark.

On the other hand, in heavy-ion collisions hadronization of heavy quarks (as well as the one of lighter partons) can occur in a rather different way. In particular the latter takes place in a hot environment, with a large abundance of light colored thermal particles nearby with which the heavy quark can recombine, neutralizing its color charge. Accounting for this recombination processes, very often modelled as a $2\to 1$ (or $3\to 1$ in the case of baryon formation) coalescence, seems fundamental in order to reproduce several experimental observations like changes in baryon-to-meson ratios and the radial and elliptic flow of final hadrons, both in the light and in the heavy-flavor sector~\cite{Fries:2008hs,Greco:2003vf,Oh:2009zj,Song:2015ykw,Cao:2019iqs,Gossiaux:2009mk}. If, at least in some kinematic domain, heavy-quark hadronization within an expanding fireball occurs via recombination with thermal particles belonging to a fluid-cell moving with a sizable fraction of the speed of light, it is possible for the daughter hadron to be produced with a larger momentum than the one of the parent quark and this contributes to describe the collective flow observed in the final hadron distributions. Furthermore, the production of baryons or strange hadrons is no longer suppressed by the necessity of exciting diquark-antiquark or $s\bar s$ pairs from the vacuum and this can change the relative abundances of the detected particles.

So far we have described hadronization as a process occurring in a completely different way in ``elementary'' (e.g. $e^+e^-$, proton-proton) and nuclear collisions. Things are actually more complex. Already in $e^+e^-$ collisions at LEP there was evidence of \emph{color-reconnection}. Consider for instance an $e^+e^-\to W^+W^-\to q_1\bar q_2 q_3\bar q_4$ event just above the $W^+W^-$ threshold. In this case the $W$ bosons are produced almost at rest and hence each of them decays into a back-to-back quark-antiquark pair ($q_1\bar q_2$ and $q_3\bar q_4$, respectively) in a color-singlet state, with a string stretched between the two color-connected partons. However, due to the peculiar kinematics, $q_1$ is often closer to $\bar q_4$ and $\bar q_2$ to $q_4$. Hence, it would be energetically more convenient to stretch a string between the quark and the antiquark \emph{closer in space} than between the ones resulting color connected from the perturbative calculation~\cite{Sjostrand:1993rb,Sjostrand:1993hi}. There is evidence that this \emph{non-perturbative} color-reconnection actually takes place~\cite{ALEPH:2013dgf}. Color reconnections are currently implemented in all multi-purpose QCD event generators and look important for the description of the underlying event in hadronic collisions; in particular, they were shown to give rise to flow-like signatures in proton-proton collisions at LHC energies~\cite{OrtizVelasquez:2013ofg}.

Actually, there are also other important observables studied in the collisions of ``elementary'' hadrons which cannot be interpreted within the factorized description of hadronization in terms of universal fragmentation fractions/functions.
An example is the so-called \emph{beam-remnant drag}, i.e. the tendency in a hadronic collision to produce at forward rapidity mainly particles which share a valence quark (or diquark) with the remnant of the projectile moving in the same direction. Focusing only on the heavy-flavour sector, asymmetries in the production of $D^-/D^+$~\cite{WA82:1993ghz,E769:1993hmj,E791:1996htn,WA89:1998wdl}, $D_s^-/D_s^+$~\cite{E791:1997eip,WA89:1998wdl}, $\Lambda_c^+/\bar\Lambda_c^-$~\cite{WA89:1998wdl,SELEX:2001iqh} and $\Lambda_b^0/\bar\Lambda_b^0$~\cite{D0:2015rnb} in various hadronic collisions have been observed. They can be interpreted assuming that, in analogy with what occurs in nuclear collisions, the heavy quark hadronize via recombination with some other parton already present in the system, in this case belonging to the beam-remnant~\cite{Norrbin:1998bw,Norrbin:2000zc}.

However, the biggest surprise came from the study of $\Lambda_c^+$ production in $pp$ collisions at the LHC~\cite{ALICE:2020wfu,ALICE:2020wla,ALICE:2021rzj}, where the ratio $\Lambda_c^+/D^0$ turned out to be a factor 5 bigger than the one measured in $e^+e^-$~\cite{Gladilin:2014tba} or $ep$ collisions~\cite{ZEUS:2010cic}. Hence, all QCD event generators with heavy-flavor fragmentation tuned on charm production in $e^+e^-$ collisions (PYTHIA 8 with Monash tune~\cite{Skands:2014pea}, HERWIG~\cite{Bellm:2015jjp}, POWHEG~\cite{Frixione:2007nw}) dramatically fail in reproducing the $\Lambda_c^+$ cross section at the LHC. A better agreement with the data is obtained with PYTHIA tunes including multiple partonic interactions (MPI), color-reconnections beyond the leading-color approximation and baryon-junction formations~\cite{Christiansen:2015yqa,Bierlich:2015rha}. An alternative approach, developed by the Catania group, assumes that also in $pp$ collisions a small QGP droplet can be formed, in which the charm quarks of not too large transverse momentum can hadronize via recombination with light thermal partons~\cite{Minissale:2020bif}. This hadronization mechanism, analogous to the one at work in nucleus-nucleus collisions, naturally explains the $p_T$ dependence of the $\Lambda_c^+/D^0$ ratio. 
The latter can be studied also for different densities of the produced system, from low to high-multiplicity $pp$ collisions~\cite{ALICE:2021npz} and from peripheral to central nucleus-nucleus collisions~\cite{ALICE:2021bib}. In both cases one observes a shift of the $\Lambda_c^+$ production towards larger values of $p_T$ when the surrounding environment gets denser, suggesting a common mechanism of hadronization in these collisions which leads $\Lambda_c^+$ to inherit the radial flow of the fireball.  

The above observations motivate us to develop an improved model of heavy-quark hadronization in high-energy nucleus-nucleus (and possibly $pp$) collisions, to interface to our POWLANG transport setup describing the stochastic propagation of the heavy quarks in the hot deconfined fireball produced in the event~\cite{Alberico:2011zy,Alberico:2013bza,Beraudo:2014boa,Beraudo:2017gxw,Beraudo:2021ont}. Actually, while in our first studies heavy quarks crossing the freeze-out hypersurface were assumed to undergo independent in-vacuum fragmentation~\cite{Alberico:2011zy,Alberico:2013bza}, in later works~\cite{Beraudo:2014boa,Beraudo:2017gxw,Beraudo:2021ont} we introduced a new description of hadronization, in which charm and beauty quarks/antiquarks are recombined with light thermal antiquarks/quarks from the medium into color-singlet strings which are then fragmented according to the Lund model. Hence, within this second scheme, the collective flow of the light thermal partons affects the kinematic distributions of the final charm and beauty hadrons, whose radial, elliptic, triangular...flow receive and important contribution from in-medium hadronization. However, this is not sufficient to modify the heavy-flavor hadrochemistry with respect to elementary collisions, since for instance within such a setup the production of a $\Lambda_c^+$ baryon would require the excitation of a diquark-antidiquark pair from the vacuum near the charm endpoint of the string, a quite rare process due to the high mass of the pair. Therefore, one would not be able to reproduce for instance the high value of the $\Lambda_c^+/D^0$ ratio observed in heavy-ion (but also $pp$) collisions.\\
In order to overcome the above shortcoming here we introduce some improvements to our picture of hadronization, for the moment focusing on the case of charm. First of all we assume that a charm quark at the hadronization hypersurface can recombine not only with a light antiquark, but also with a diquark, assuming several diquark states of different spin, isospin and strangeness to be present in the medium around the QCD pseudocritical temperature, each with its corresponding thermal abundance. Furthermore, we provide a special treatment of low-mass color-singlet clusters, which are decayed assuming that the daughter charm hadron inherits the flavour and baryon number of the ancestor. This is sufficient to get relative yields of the different charmed hadrons in better agreement with the experimental findings, still obtaining momentum and angular distributions which keep track of the collective flow of the medium, as in our previous hadronization scheme based on in-medium string formation.

Our paper is organized as follows. In Sec.~\ref{sec:model} we present in detail our model, illustrating our implementation of in-medium cluster formation and decay and emphasizing some very general features of such a picture of hadronization, also in connection with previous literature describing charm production in different kinds of hadronic collisions. In Sec.~\ref{sec:results} we display the results obtained with our new hadronization scheme concerning the relative yields of the different charmed hadrons, with their $p_T$-dependence, and the flow coefficients of the various heavy-flavour species. Finally, in Sec.~\ref{sec:discussion} we discuss our major findings, with the perspective of providing a similar picture also for the proton-proton case, hence studying how a hot expanding environment, in spite of its limited size, can affect charm production also in these collisions. Details on the modelling of the background medium and on the simulation of the heavy-quark transport in the deconfined fireball produced in nuclear collisions can be found in Appendices~\ref{App:hydro} and~\ref{App:transport}.

\section{New hadronization model}\label{sec:model}
\begin{table}
   \begin{center}  
  \begin{tabular}{|c|c|c|c|c|}
\hline
Species & $g_s$ & $g_I$ & $M$ (GeV) & daughter (if $M_{\cal C}\!<\!M_{\rm max}$)\\
\hline
$l$ & 2 & 2 & 0.33000 & $D^0,D^+$\\
\hline
$s$ & 2 & 1 & 0.50000 & $D_s^+$\\
\hline
$(ud)_0$ & 1 & 1 & 0.57933 & $\Lambda_c^+$\\
\hline
$(ll)_1$ & 3 & 3 & 0.77133 & $\Lambda_c^+$\\
\hline
$(sl)_0$ & 1 & 2 & 0.80473 & $\Xi_c^0,\Xi_c^+$\\
\hline
$(sl)_1$ & 3 & 2 & 0.92953 & $\Xi_c^0,\Xi_c^+$\\
\hline
$(ss)_1$ & 3 & 1 & 1.09361 & $\Omega_c^0,\Xi_c^+$\\
\hline
   \end{tabular}
       \end{center}
\caption{The light thermal particles involved in the recombination process, with their spin and isospin degeneracy, their mass and the daughter charmed hadrons arising from the cluster decay. In the above $l=u,d$, assuming isospin symmetry.}\label{tab:masses}
\end{table}
\begin{figure*}[!h]
  \begin{center}
\includegraphics[clip,width=0.45\textwidth]{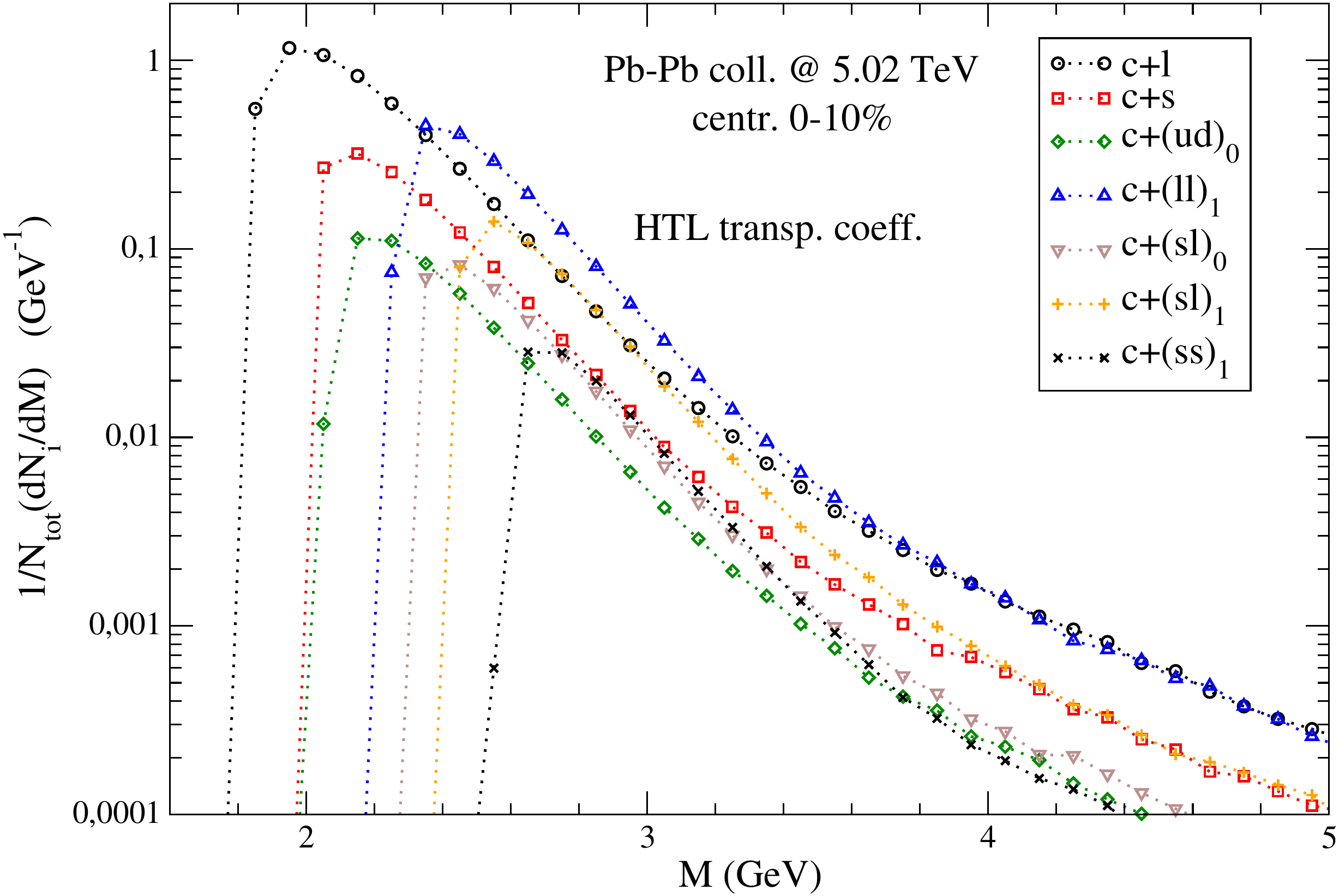}
\includegraphics[clip,width=0.45\textwidth]{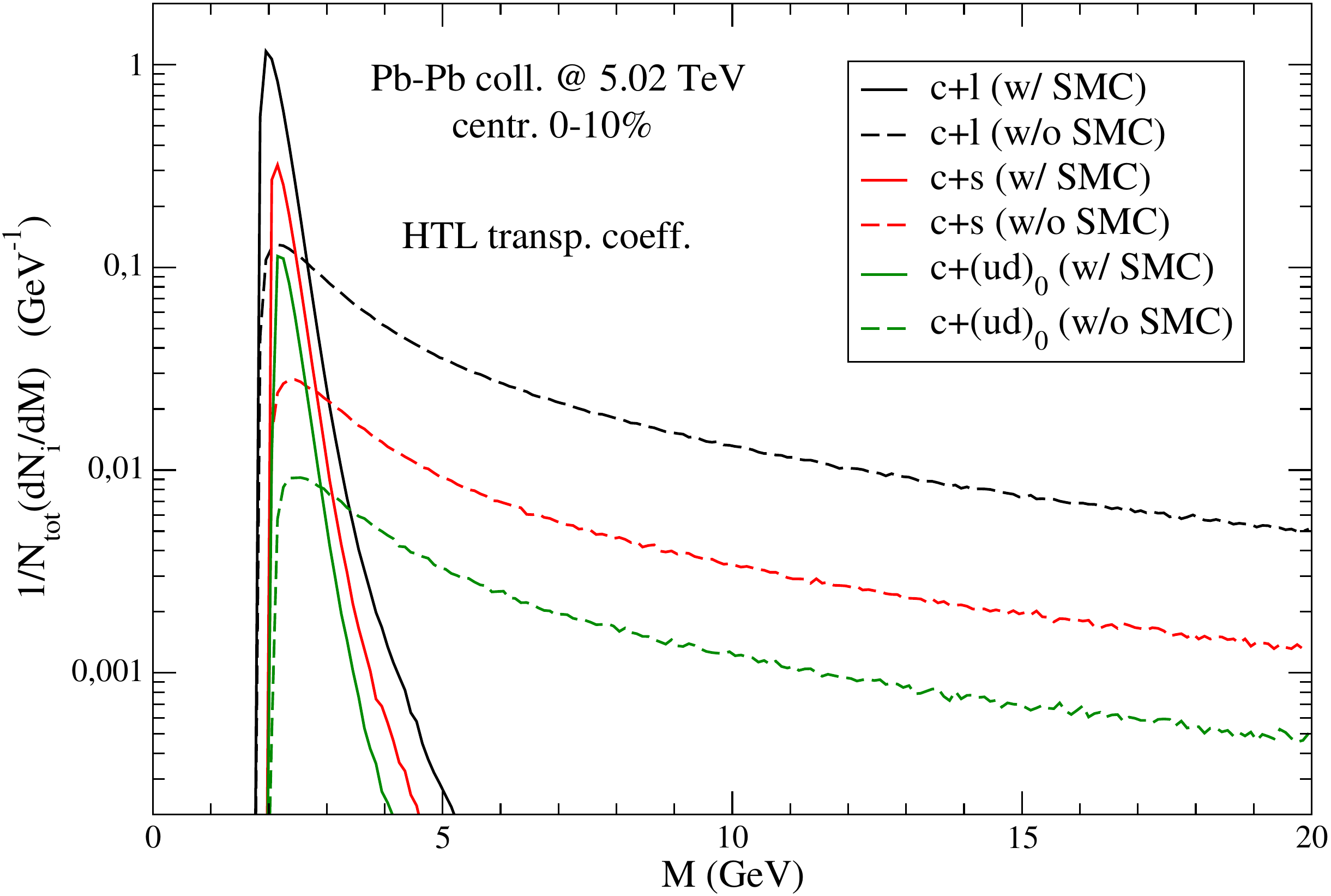}
\caption{Left panel: the invariant-mass distribution of the $Q\overline q$ and $Q(qq)$ clusters at the hadronization hypersurface in the different isospin, strangeness and spin channels. Right panel: cluster invariant-mass distributions in case recombination occurs locally (continuous curves, default implementation), with strong correlation between HQ momentum and position, or non-locally (dashed curves), with no space-momentum correlation. All results refer to central Pb-Pb collisions at $\sqrt{s_{\rm NN}}\!=\!5.02$ TeV, with charm quark propagation before hadronization described by HTL transport coefficients.}\label{fig:Mdistr}
\end{center}
\end{figure*}
\begin{figure*}[!h]
    \begin{center}
\includegraphics[clip,width=0.45\textwidth]{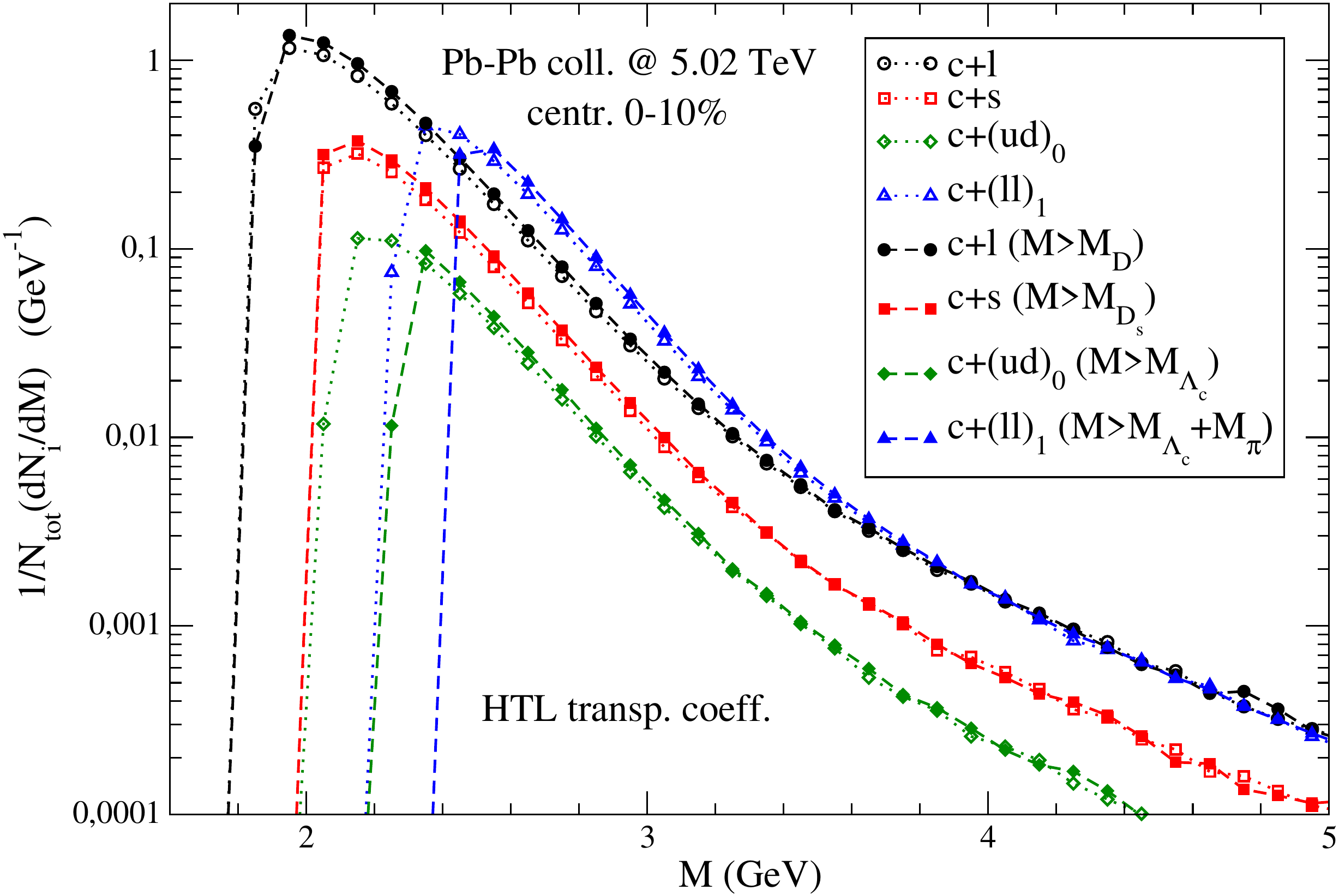}
\includegraphics[clip,width=0.45\textwidth]{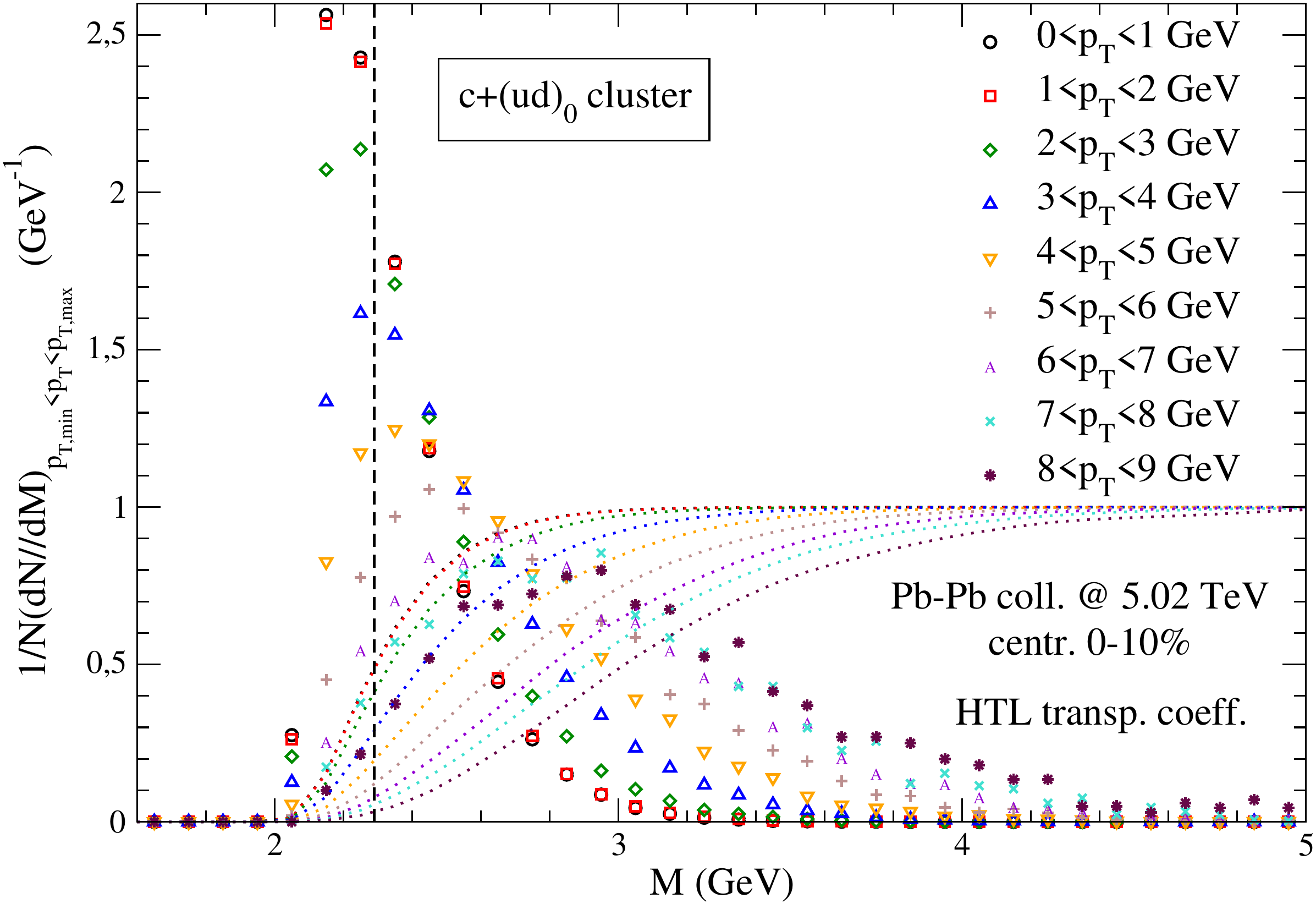}
    \caption{Left panel: invariant-mass distributions of charm clusters $\cal C$ before (open symbols) and after (filled symbols) resampling in order to allow the two-body decay ${\cal C}\to h_c+\gamma/\pi$. Right panel: invariant-mass distributions (normalized to 1) of $c(ud)_0$ clusters for different $p_T$-ranges of the parent charm quark (symbols), together with their cumulative integrals (dotted curves) and the $\Lambda_c^+$ threshold (vertical dashed line).}\label{fig:Mdistr-cuts}
    \end{center}  
\end{figure*}
We now present in detail our new hadronization model, here applied to the study of charm production in heavy-ion collisions after the simulation of the heavy-quark transport in the hot deconfined fireball performed through a relativistic Langevin equation. Actually, the algorithm we developed is pretty general and simply requires the knowledge of the position and momentum components of the heavy quarks on the hadronization hypersurface, together with the velocity of the corresponding fuid cells, independently from how the previous evolution is simulated.

First of all one defines a hadronization hypersurface, identified by the temperature $T_H$, here set equal to 155 MeV in agreement with lattice-QCD estimates of the pseudocritical temperature (we neglect a possible flavor dependence of the deconfinement temperature~\cite{Bellwied:2013cta}). Once a charm quark, during its stochastic propagation in the fireball, reaches a fluid-cell belonging to such a hypersurface it is recombined with a thermal particle from the same cell. The latter can be either a light antiquark ($u,d,s$) or -- at variance with our previous hadronization setup-- a diquark, assuming that around the QCD pseudocritical temperature several diquark states exist in the medium. For their masses, summarized in Table~\ref{tab:masses}, we take the default values employed by PYTHIA 6.4~\cite{Sjostrand:2006za}. Light quarks and diquarks are assumed to be thermally distributed in the local rest frame of the fluid cell.
Our algorithm consists in the following steps:
\begin{enumerate}
\item As a first step we randomly extract the medium particle species with which the charm quark undergoes recombination, with a statistical weight given by its corresponding thermal density at the temperature $T_H$, including its spin $g_s$ and isospin $g_I$ degeneracy:
\beq 
n=g_s\,g_I\,\frac{T_H M^2}{2\pi^2}\sum_{n=1}^{\infty}\frac{(\mp 1)^{n+1}}{n}\,K_2\left(\frac{nM}{T_H}\right)\,.
\eeq 
In the above $M$ is the quark (diquark) mass and the $-/+$ sign refers to fermions/bosons, respectively, even if the effect of quantum statistics turns out to be numerically negligible and for all medium particles one can keep only the $n=1$ term in the expansion, which corresponds to the Boltzmann limit;
\item Once selected the medium particle, we extract its three-momentum in the local rest frame of the fluid-cell from a thermal distribution;
\item We boost the four-momentum of the medium particle to the laboratory frame and recombine the latter with the heavy quark, constructing the cluster $\cal C$;
\item We evaluate the invariant mass $M_{\cal C}$ of the cluster. If the latter is smaller than the mass of the lightest charmed hadron in that channel we go back to point 1 and extract a new particle. Otherwise we continue and go to point 5;
\item After introducing an intermediate cutoff $M_{\rm max}$ we simulate the decay of the cluster, depending on its invariant mass:
\begin{enumerate}
\item If the cluster is sufficiently light ($M_{\cal C}<M_{\rm max}$) we simulate its isotropic two-body decay in its own rest frame. The daughter charmed hadron will carry the baryon number and strangeness of the parent cluster. The companion light particle will be a pion (in most of the cases) or a ``photon'', to be interpreted either as a real particle if this last decay channel is predicted by the Particle Data Group~\cite{ParticleDataGroup:2020ssz} or just as a trick to ensure four-momentum conservation in the decay, if the invariant mass of the cluster is so low to forbid pion production. Notice that in PYTHIA there is a similar problem in the simulation of the decay of very light clusters and one also conserves four-momentum in these ``cluster-collapse'' processes through the exchange with the medium of fictitious particles called ``soft gluons''~\cite{Norrbin:2000zc}. The only exception to the above picture is represented by the $c(ss)_1$ cluster, for which the decay ${\cal C}\to \Xi_c^+ K^-$ is predicted by the PDG~\cite{ParticleDataGroup:2020ssz} and inserted in our algorithm, with one unit of strangeness being carried away by the kaon.
\item If the cluster is heavier than the above cutoff ($M_{\cal C}>M_{\rm max}$) the latter is fragmented as a Lund string as in our previous hadronization scheme, employing the string-fragmentation routine provided by PYTHIA 6.4~\cite{Sjostrand:2006za}. 
\end{enumerate}
Since low/high-momentum charm quarks mainly give rise to light/heavy color-singlet clusters, the above mechanism allows a smooth transition from a low-$p_T$ regime, in which recombination leaves its fingerprints in the final charmed-hadron yields and kinematic distributions, to a high-$p_T$ domain, in which hadronization occurs via string fragmentation, as in the vacuum. In fact, if $M_C$ is very large, the production of the daughter charmed hadron is not very sensitive to the second endpoint of the string and the fact that in a heavy-ion collision the latter is taken from the background fireball does not play a big role.
\end{enumerate}
Reproducing the separate fragmentation fractions into $D^0$ and $D^+$ mesons requires some special care. In fact, a sizable fraction of the observed $D$ mesons arises from the strong decays of the $D^*$ resonances, which more frequently lead to the production of a $D^0$ than a $D^+$. This explains the final higher yields of $D^0$ mesons. In our model we do not include explicitly $D^*$ resonances as possible hadronization products. However when we have a $c\bar u$ or $c\bar d$ cluster in the vector channel (which occurs with probability 0.75 according to the decomposition $2\otimes 2 = 1\oplus 3$), for a broad range of values of its mass (up to $M_{\cal C}=2.8$ GeV), we simulate the two-body decay of the latter taking the $D^*$ branching ratios provided by the PDG~\cite{ParticleDataGroup:2020ssz}. As shown in the next section, this is important in order to obtain a $D^+/D^0$ ratio in agreement with the experimental data.   

In the left panel of Fig.~\ref{fig:Mdistr} we show the invariant-mass distributions of all the possible charmed clusters formed on the hadronization hypersurface of a central Pb-Pb collision at $\sqrt{s_{\rm NN}}\!=\!5.02$ TeV, depending on the light antiquark or diquark with which recombination occurs. As one can see, one obtains a steeply falling spectrum, with most of the clusters having a mass below 4 GeV. This feature is also present in an event generator like HERWIG~\cite{Webber:1983if,Bellm:2015jjp}, due to the effective ``pre-confinement'' mechanism entailed by its angular-ordered parton shower, in which collinear splittings favour the formation of low-mass color-singlet clusters. Here the origin of the rapid drop of the cluster mass distribution is different and is due to the strong correlation at hadronization among the momentum of the heavy quark, the fluid-cell it occupies and the four-velocity of the latter. This space-momentum correlation (SMC) favours configurations in which the momenta of the charm quark and of the light particle from the medium are quite collinear, leading to the formation of low-mass clusters. This can be appreciated considering the right panel of Fig.~\ref{fig:Mdistr}, in which we compare the cluster mass distributions with and without SMC's, the last case being obtained through an event-mixing approach in which the heavy quark momentum is taken from event $i$ and its position on the hadronization hypersurface is taken from a different event $j$ (here with ``event'' we mean the production of a single $Q\overline Q$ pair in the same background fireball common to all pairs). As one can see, in the absence of SMC's all distributions are characterized by long tails extending up to very high values of the cluster mass. As anticipated, such very massive clusters -- very rare in our model in which SMC's are present by construction -- would be fragmented as Lund strings giving rise to the production of a large number of hadrons. How this would affect the final relative yields and kinematic distributions of the different charmed mesons and baryons is studied in the next section. Here we wish to notice that our hadronization model, in which parton recombination occurs \emph{locally} entailing a strong SMC and leading to the formation of quite light color-singlet clusters, can be considered as an extreme version of the color-reconnection mechanism usually implemented in QCD event generator, based on the idea of minimizing the length of the color strings connecting opposite-charge partons.  

As anticipated, in a certain fraction of cases recombination does not lead to the formation of a cluster with large enough mass to allow a two-body decay, not even via the emission of a massless particle. In this case the event is resampled, extracting a new particle from the medium with which the heavy quark is recombined. Whether it is better to extract a new particle or to simply resample its momentum is a matter of choice. Here we choose the first option, knowing that in any case this represents a source of systematic uncertainties for the yields and kinematic distributions of the final charmed hadrons.
In order to quantify such an uncertainty, for a few cases, we also display results corresponding to a resampling of the momentum while keeping the particle species fixed. In this case the final yields of charmed hadrons are fixed by the equilibrium abundance of the medium particles involved in the recombination process.

How often resampling is required strongly depends on the values of the quark and diquark masses, knowing that one is dealing with effective masses, which do not necessarily coincide with the ones employed in the generation of the hard event (the $Q\overline Q$ pair in this case). Concerning the light-quark and diquark sector we keep the values given in Table~\ref{tab:masses} fixed. For the charm quark we explore the values $m_c=1.3$ GeV and $m_c=1.5$ GeV. On average, independently from the centrality class, hadronizing 20 millions charm quarks requires sampling about 45 millions light antiquarks or diquarks from the medium if $m_c=1.3$ GeV and about 23 millions if $m_c=1.5$ GeV. In order to minimize the systematic uncertainties arising from the resampling procedure, our final results will refer to this second value. 
In the left panel of Fig.~\ref{fig:Mdistr-cuts} we display the cluster mass distributions in the different channels before and after resampling. The results refer to the $m_c=1.5$ GeV case. As can be seen, resampling mainly involves clusters in which a charm quark is recombined with a light diquark, which in a non-negligible fraction of events are below the threshold allowing a two-body decay. Overall, our resampling procedure leads to a slight decrease of the production of charmed baryons and a corresponding slight increase of the one of charmed mesons with respect to the yields expected from the thermal abundance of light quarks and diquarks. Charmed-baryon relative yields are in any case much larger than the one predicted by fragmentation fractions tuned to $e^+e^-$ collisions. In the right panel of Fig.~\ref{fig:Mdistr-cuts} we focus on one of the channels leading to the production of $\Lambda_c^+$ baryons and we display the cluster mass distribution for different values of the transverse momentum of the parent charm quark, together with its cumulative integral. Comparing the resulting curves with the location of the $\Lambda_c^+$ threshold one can see that the resampling mainly affects baryonic clusters arising from a low-$p_T$ charm quark.

\section{Results}\label{sec:results}
\begin{figure*}[!h]
  \begin{center}
    \includegraphics[clip,width=0.95\textwidth]{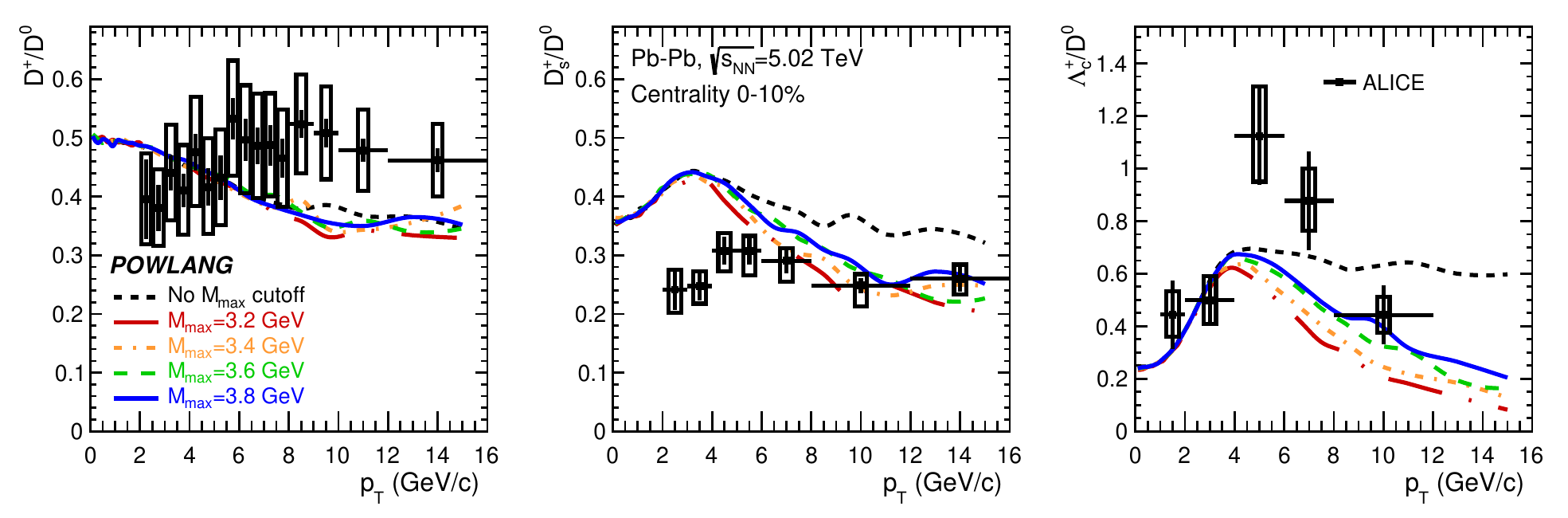}
\caption{Dependence of the charmed-hadron ratios on the critical cluster mass $M_{\rm max}$ above which the latter is fragmented as a Lund string rather than suffering a two-body decay. Theoretical results are compared to experimental data obtained in central Pb-Pb collisions at $\sqrt{s_{\rm NN}}\!=\!5.02$ TeV~\cite{ALICE:2021bib,ALICE:2021kfc,ALICE:2021rxa}. In the following we choose the value $M_{\rm max}=3.8$ GeV.}\label{Fig:Mmax}
    \end{center}
  \end{figure*}
\begin{figure*}[!h]
    \begin{center}
      \includegraphics[clip,width=0.95\textwidth]{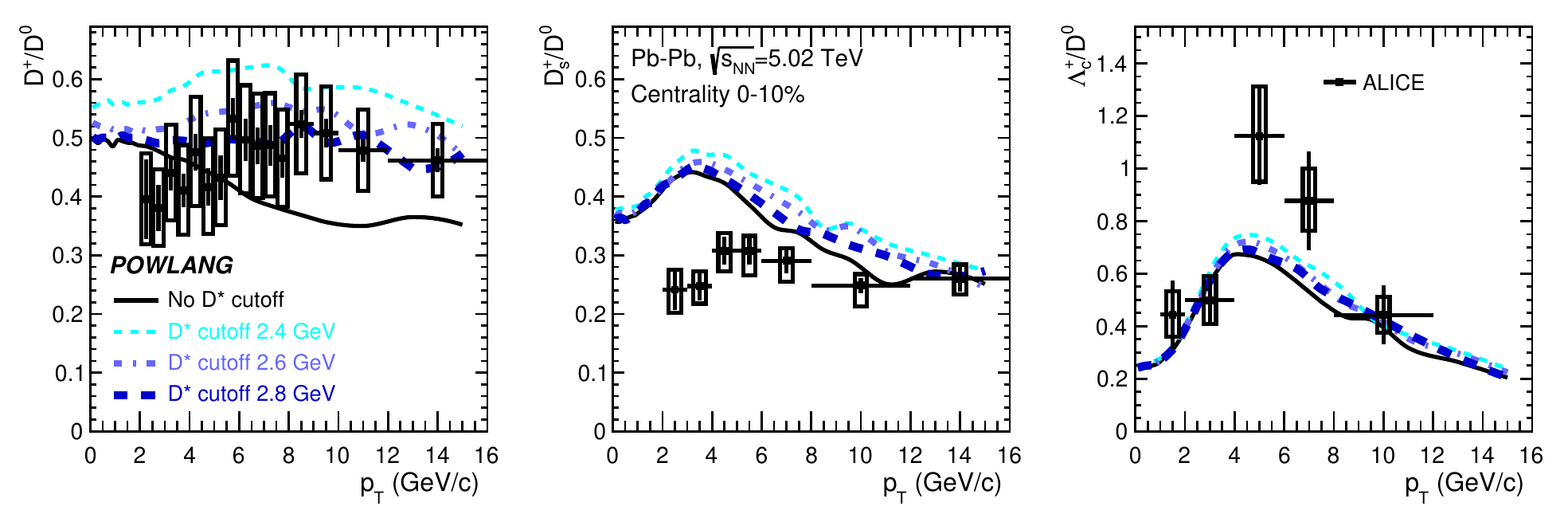}
\caption{Sensitivity of the relative charmed-hadron yields on the decay branching ratios of $c\overline l$ clusters of spin 1. The $D^{*0}$ and $D^{*+}$ branching ratios are employed for different ranges of the cluster mass $M_{\cal C}$. The upper bound $M_{\cal C}<2.8$ GeV is employed in the following. Theoretical results are compared to experimental data obtained in central Pb-Pb collisions at $\sqrt{s_{\rm NN}}\!=\!5.02$ TeV~\cite{ALICE:2021bib,ALICE:2021kfc,ALICE:2021rxa}.}\label{Fig:D*}
        \end{center}
\end{figure*}
In this section we present and discuss the results for the charmed-hadron yields and flow coefficients in nucleus nucleus collisions obtained interfacing our new hadronization algorithm to the output of transport calculations describing the heavy-quark propagation in the deconfined fireball.
As usual, these simulations are based on the relativistic Langevin equation. Its numerical implementation as well as the modelling of the initialization and hydrodynamic evolution of the background medium are discussed in detail in our previous publications~\cite{Alberico:2011zy,Alberico:2013bza,Beraudo:2014boa,Beraudo:2017gxw,Beraudo:2021ont} and are briefly summarized in Appendices~\ref{App:hydro} and~\ref{App:transport}.

Before displaying the predictions obtained with our new hadronization scheme, we discuss how its few free parameters have been fixed through a comparison with experimental data obtained in central Pb-Pb collisions at $\sqrt{s_{\rm NN}}\!=\!5.02$ TeV~\cite{ALICE:2021bib,ALICE:2021kfc,ALICE:2021rxa}.
In Fig.~\ref{Fig:Mmax} we consider the effect of the parameter $M_{\rm max}$, which regulates the transition from the invariant-mass region in which a two-body cluster decay is performed to the one in which hadrons are produced through a $N$-body string fragmentation. In the plots the black curves (labelled as ``No $M_{\rm max}$ cutoff'') are obtained performing a two-body decay for all charmed clusters, independently from their invariant mass. The colored curves are instead obtained assuming that a cluster undergoes an isotropic two-body decay only if its mass $M_{\cal C}$ is lower than $M_{\rm max}$; heavier clusters, with an invariant mass $M_{\cal C}>M_{\rm max}$, are fragmented as strings according to the Lund model implemented in PYTHIA 6.4~\cite{Sjostrand:2006za}. As can be seen, forcing a two-body decay of the clusters independently from their mass does not allow one to reproduce the qualitative behavior of the experimental data: one would for instance obtain a flat $\Lambda_c^+/D^0$ ratio as a function of $p_T$, while experimental data indicate that the latter tends to decrease for large transverse momenta, approaching the value around 0.1 found in $e^+e^-$ collisions. The transition to a string-fragmentation picture for the hadronization of high-mass clusters provides a better agreement with the data. In fact, the fraction of high-mass clusters ($M_{\cal C}>M_{\rm max}$) increases with the momentum of the parent charm quark, as already seen in Fig.~\ref{fig:Mdistr-cuts}. These color-singlet objects are then fragmented as standard Lund strings, in analogy to what done in elementary collisions, which explains the agreement with the $e^+e^-$ result obtained at high $p_T$. Concerning the value of $M_{\rm max}$, in the following we set $M_{\rm max}\!=\!3.8$ GeV, which turns out to be very close to the one around 4 GeV employed in HERWIG, where light clusters undergo an isotropic two-body decay, while heavier ones first suffer a fission into two lower-mass clusters. Notice that the two hadronization mechanisms differ not only for the number of produced particles, but also for their angular distribution. In fact, the two-body decay of a low-mass cluster is assumed to occur isotropically in its local rest frame; on the contrary, in the fragmentation of a string, particles are emitted preferentially along the longitudinal direction with respect to its yo-yo motion. In PYTHIA a smooth transition between the two regimes is introduced, which is not included in the present implementation of our model.   

In high-energy collisions a sizable fraction of $D$-meson production comes from the decay of $D^{*0}$ and $D^{*+}$ resonances. This affects the relative yields of $D^0$ and $D^+$ mesons; in particular, since $D^*$ resonances tend to decay preferentially into a $D^0$ meson, one gets $D^+/D^0<1$, at variance with what naively expected from isospin symmetry. 
In our model we do not produce directly $D^*$ resonances, however -- for a broad range of invariant masses -- in simulating the two-body decay of $c\overline q$ clusters in the spin-1 channel we employ the $D^*$ branching ratios provided by the PDG. In Fig.~\ref{Fig:D*} we display how our predictions are sensitive to the above range. As one can see, taking the $D^*$ branching ratios for all the two-body decays of vector clusters (black curve) leads to a decreasing behavior of the $D^+/D^0$ ratio as a function of $p_T$ not supported by the data. On the other hand employing the $D^*$ decay channels for a too narrow range of invariant masses leads to overpredict the $D^+/D^0$ ratio. We found a good compromise employing the $D^*$ branching ratios up to a cluster mass $M_{\cal C}\!=\!2.8$ GeV. In the following, our results refer to this choice.  

\begin{figure*}[!h]
      \begin{center}
        \includegraphics[clip,width=0.95\textwidth]{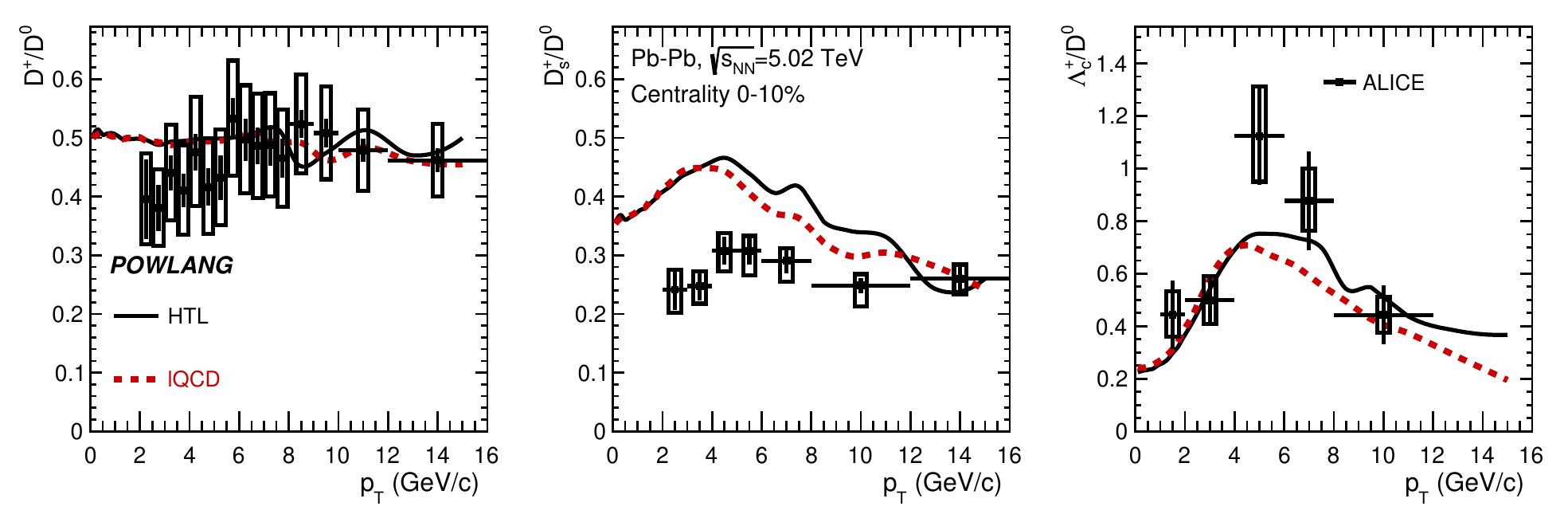}
\caption{Predictions for the relative yields of charmed hadrons (relative to $D^0$ mesons) in central Pb-Pb collisions at $\sqrt{s_{\rm NN}}\!=\!5.02$ TeV for different transport coefficients compared to recent ALICE data~\cite{ALICE:2021bib,ALICE:2021kfc,ALICE:2021rxa}.}\label{Fig:D-Ds-Lc}
        \end{center}
\end{figure*}
\begin{figure*}[!h]
      \begin{center}
        \includegraphics[clip,width=0.95\textwidth]{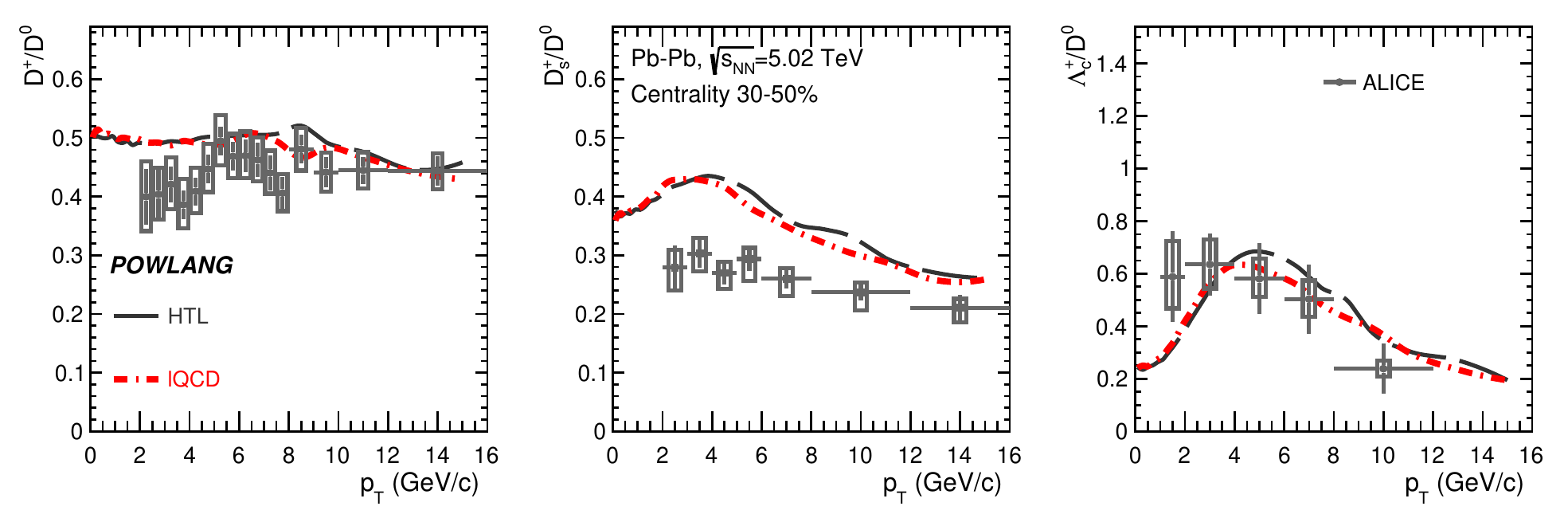}
\caption{The same as in Fig.~\ref{Fig:D-Ds-Lc}, but for semi-central Pb-Pb collisions.}\label{Fig:D-Ds-Lc-NC}
        \end{center}
\end{figure*}
After fixing the few parameters of our hadronization model, we now move to compare its predictions for charmed hadron production in nuclear collisions to the currently available experimental data. In particular, in Figs.~\ref{Fig:D-Ds-Lc} and~\ref{Fig:D-Ds-Lc-NC} we display our results for the $D^+/D^0$, $D_s^+/D^0$ and $\Lambda_c^+/D^0$ ratios as a function of $p_T$ in central (0-10\%) and semi-central (30-50\%) Pb-Pb collisions at $\sqrt{s_{\rm NN}}\!=\!5.02$ TeV for different heavy-quark transport coefficients, either evaluated in a weak-coupling calculation (HTL curves) or extracted from lattice-QCD simulations (lQCD curves). Differences in the theoretical results arising from the modelling of heavy-quark transport in the deconfined fireball are quite small, although HTL transport coefficients tend to give slightly higher values of the $D_s^+/D^0$ and $\Lambda_c^+/D^0$ ratios at moderate-large $p_T$.\\
Theoretical results are compared to recent experimental data obtained by the ALICE collaboration~\cite{ALICE:2021bib,ALICE:2021kfc,ALICE:2021rxa}. The major feature that one can observe, both in the experimental data and in the model prediction, is the strong enhancement of the relative production of $\Lambda_c^+$ baryons, with a $\Lambda_c^+/D^0$ ratio much larger than the value around 0.1 expected from $e^+e^-$ collisions. Experimental data display a marked dependence on the collision centality, with a sharp peak in the $\Lambda_c^+/D^0$ ratio at moderate $p_T$ in the 0-10\% most central events, which tends to disappear in more peripheral collisions. This feature can be only partially reproduced by our model, whose results for this observable display a milder centrality dependence.\\
Concerning the $D_s^+/D^0$ ratio in Pb-Pb collisions, experimental data indicate that the latter is enhanced with respect to the $pp$ reference~\cite{ALICE:2021kfc}, in particular for values of $p_T$ around $4-6$ GeV. In our model, as also shown later on while discussing the charm fragmentation fractions, we also find an enhanced production of $D_s^+$ meson in Pb-Pb collisions at LHC center-of-mass energies, but we tend to overshoot the experimental data for the $D_s^+/D^0$ ratio, in particular for low transverse momenta.    

\begin{figure*}[!h]
      \begin{center}
\includegraphics[clip,width=0.45\textwidth]{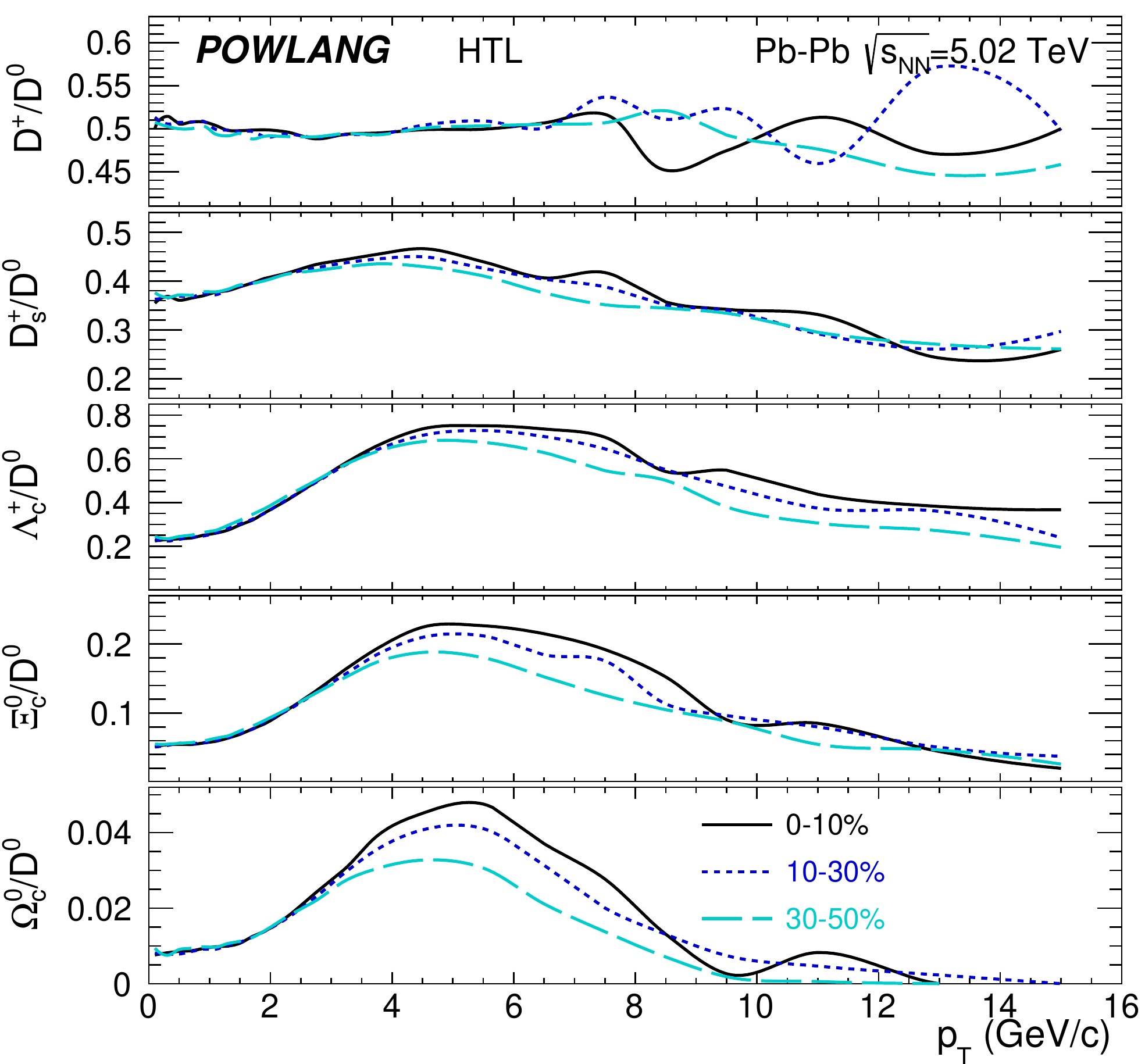}
\includegraphics[clip,width=0.45\textwidth]{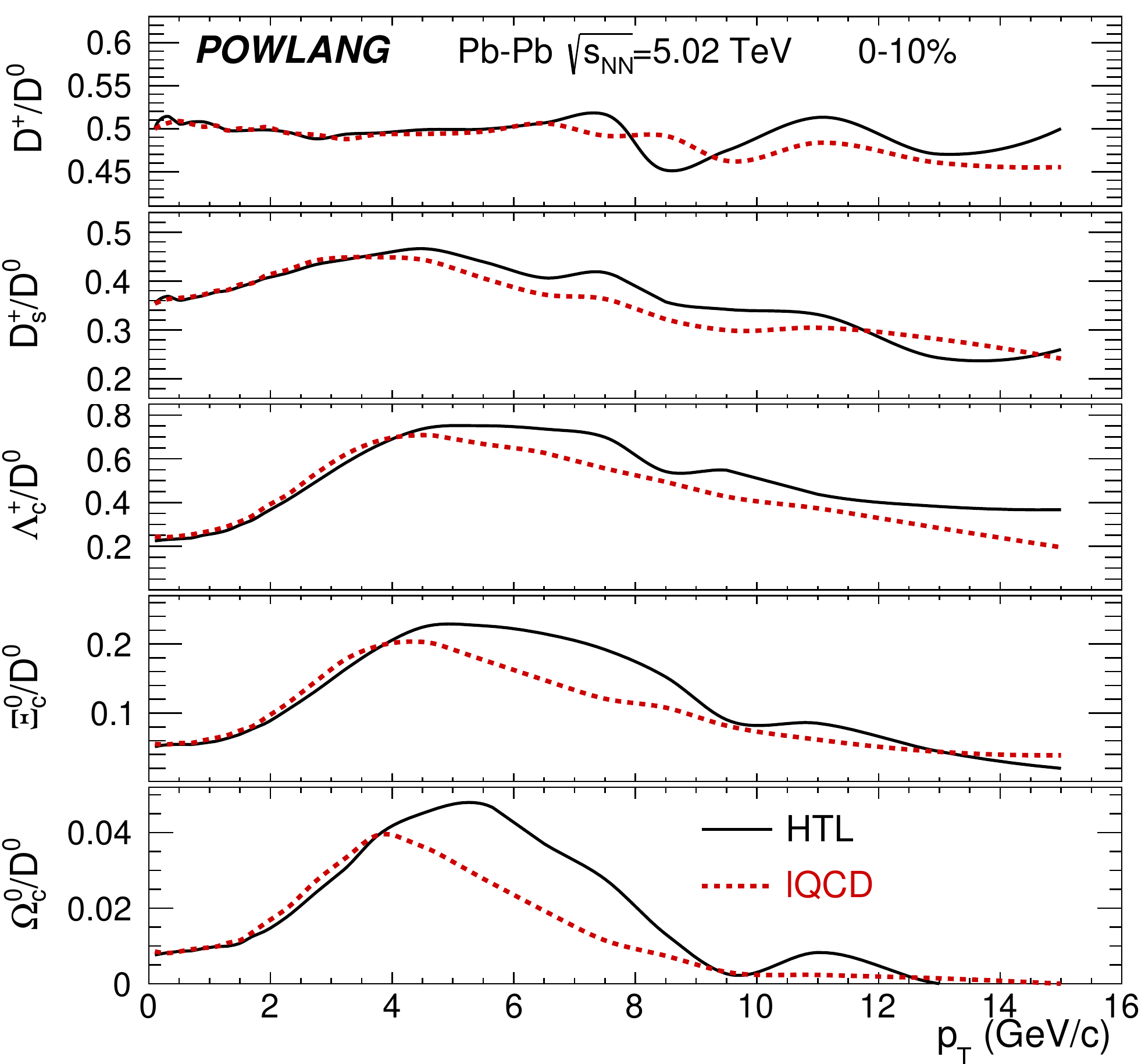}
\caption{Dependence of the relative yields of charmed hadrons as a function of $p_T$ in  Pb-Pb collisions at $\sqrt{s_{\rm NN}}\!=\!5.02$ TeV on the centrality class (left panel) and on the transport coefficients (right panel).}\label{Fig:ratios-full}
    \end{center}
\end{figure*}
\begin{figure*}[!h]
      \begin{center}
\includegraphics[clip,width=0.45\textwidth]{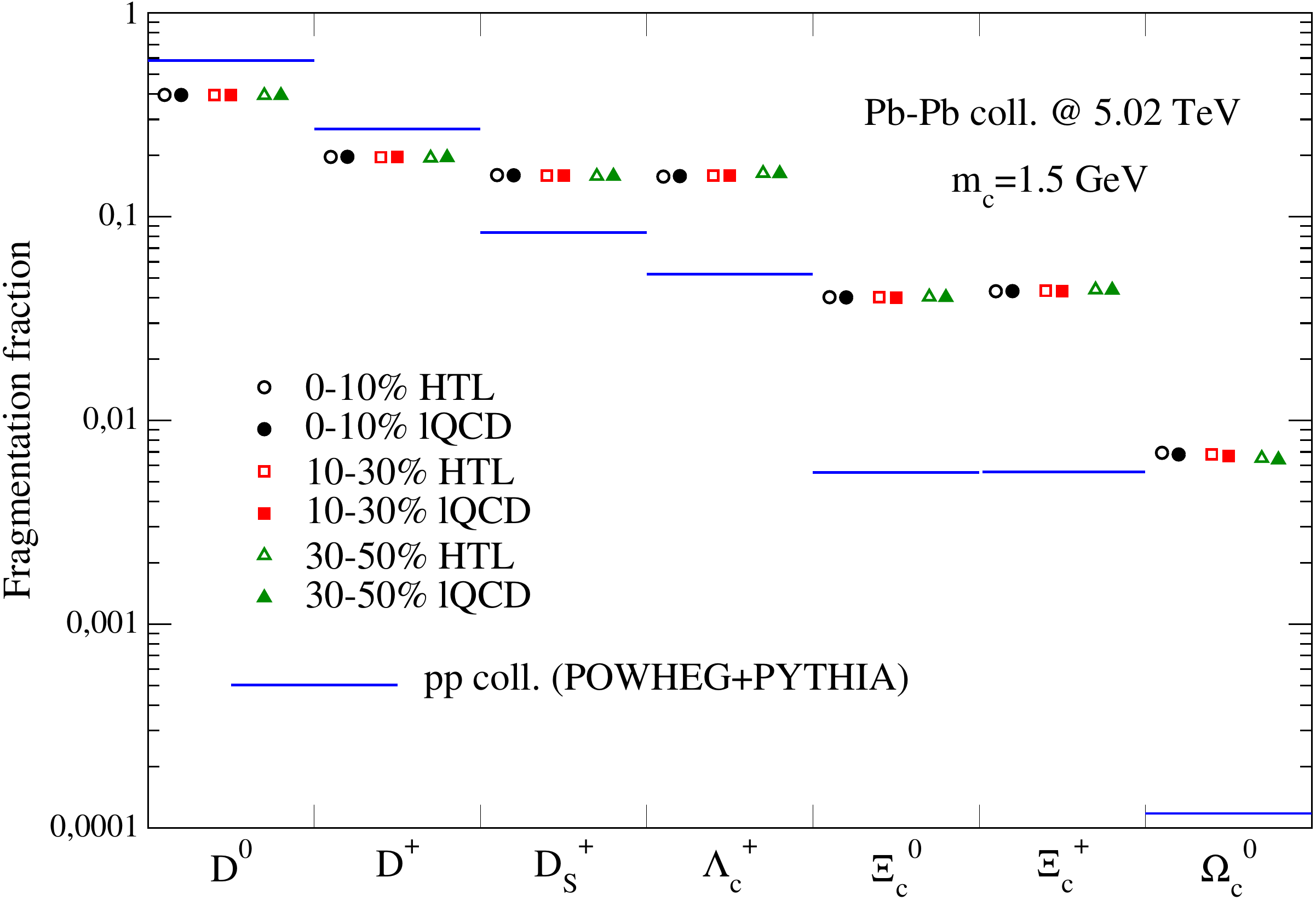}
\includegraphics[clip,width=0.45\textwidth]{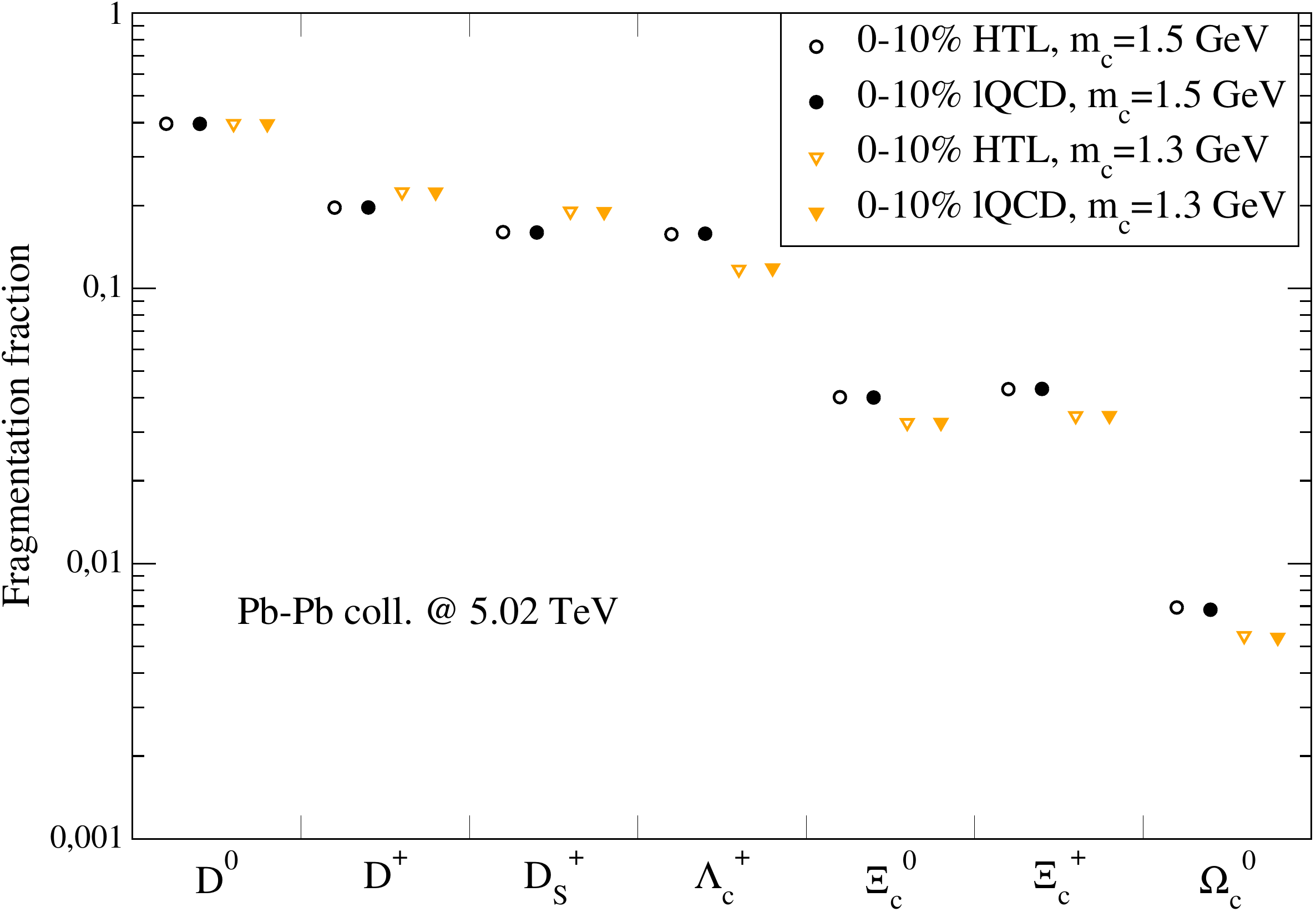}
\caption{Charm fragmentation fractions in Pb-Pb collisions for different centrality classes, transport coefficients (left panel) and values of the effective quark mass at hadronization (right panel). In the left panel we also show, in blue, the corresponding results obtained simulating proton-proton collisions at the same center-of-mass energy with the POWHEG-BOX package~\cite{Alioli:2010xd}.}\label{Fig:FF}
    \end{center}
\end{figure*}

In Fig.~\ref{Fig:ratios-full} we show the complete set of our results, including also our predictions for the production of $\Xi_c^0$ and $\Omega_c^0$ baryons as a function of $p_T$ and studying how all the ratios vary depending on the centrality of the collision (left panel) and on the choice of the transport coefficients (right panel). In order to interpret the results it is useful to study this figure together with the left panel of Fig.~\ref{Fig:FF}, where we plot the fragmentation fractions into the different charmed hadrons obtained from the $p_T$-integrated distributions. As one can see, the fragmentation fractions turn out to be rather independent both from the collision centrality and from the choice of transport coefficients. Hence, the differences visible in Fig.~\ref{Fig:ratios-full} have to be interpreted as a reshuffling of the hadrons in the different $p_T$-bins which does not affect their $p_T$-integrated relative yields. In particular, moving from peripheral to central collisions, the effect of the additional radial flow acquired via recombination looks more and more relevant, especially for the heaviest charmed hadrons, resulting in a more pronounced bump in the baryon/meson ratios at intermediate $p_T$. Differences between the HTL and lattice-QCD curves have to be attributed to the different slope of the parent charm-quark $p_T$ distributions at the end of their propagation in the deconfined fireball. On the contrary, the $D^+/D^0$ ratio looks pretty independent from the centrality of the collision, since in this case the charm quark is recombined either with a $\overline u$ or $\overline d$ antiquark which, having the same mass, are equally affected by the collective flow of the fireball.

\begin{figure*}[!h]
      \begin{center}
        \includegraphics[clip,width=0.95\textwidth]{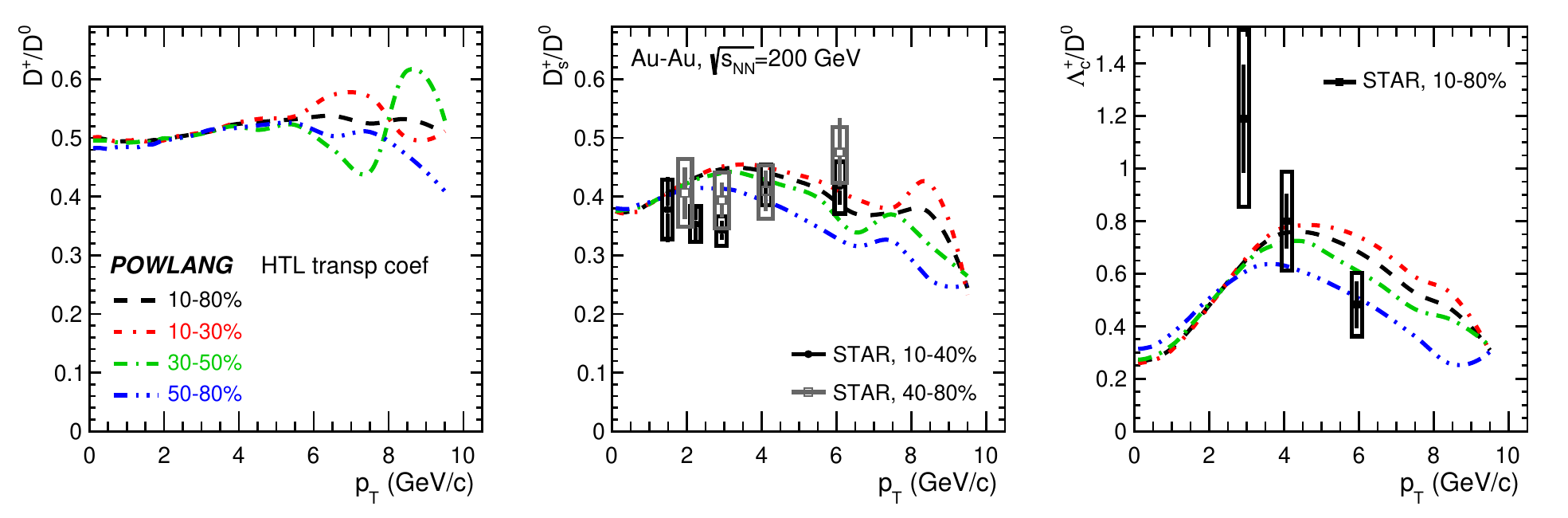}
\caption{Predictions for the relative yields of charmed hadrons (relative to $D^0$ mesons) as a function of $p_T$ in Au-Au collisions at $\sqrt{s_{\rm NN}}\!=\!200$ GeV for different centrality classes compared to STAR data~\cite{STAR:2021tte,STAR:2019ank}.}\label{Fig:cH-RHIC}
        \end{center}
\end{figure*}
In order to study how our predictions vary as a function of the center of mass of the collision, in Fig.~\ref{Fig:cH-RHIC} we compare our findings for the case of Au-Au collisions at $\sqrt{s_{\rm NN}}\!=\!200$ GeV to experimental data for the $D^+/D^0$, $D_s^+/D^0$ and the $\Lambda_c^+/D^0$ ratios provided by the STAR collaboration~\cite{STAR:2021tte,STAR:2019ank}. Results obtained with weak-coupling transport coefficients for different centrality classes are displayed. As already found at LHC energies, the effect of the radial flow is manifest in the $\Lambda_c^+/D^0$ ratio, whose peak moves to higher values of $p_T$ as one selects more central events. As already observed in central collisions at LHC energies, our bump in the $\Lambda_c^+/D^0$ ratio is not as sharp as the one found in the experimental data, although still affected by large statistical and systematic uncertainties. We notice that, at RHIC energies, our model displays a better agreement with the experimental data for the $D_s^+/D^0$ ratio than what found in Pb-Pb collisions at the LHC.

\begin{figure}[!h]
      \begin{center}
\includegraphics[clip,width=0.45\textwidth]{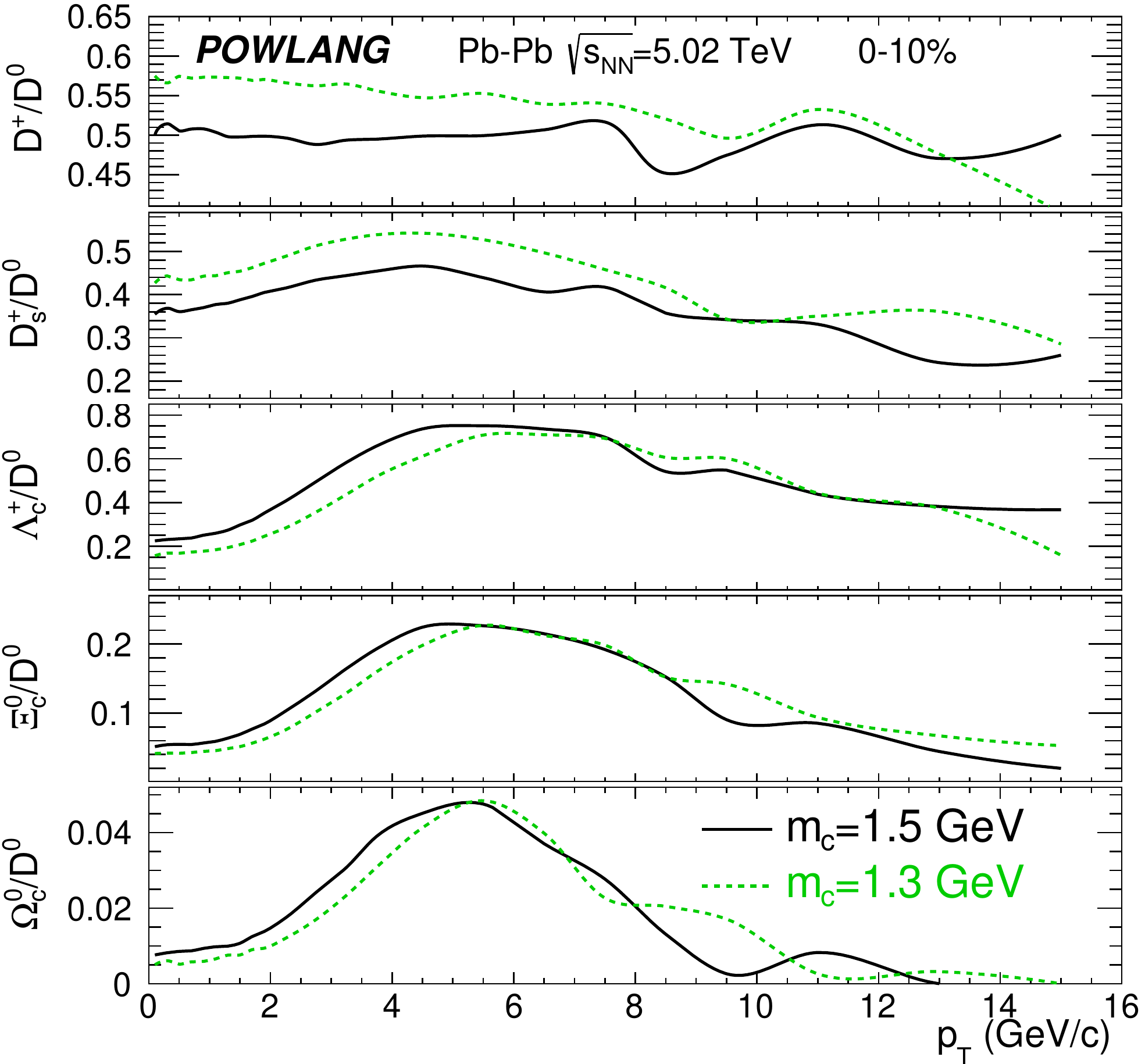}
\caption{Dependence of the relative yields of charmed hadrons as a function of $p_T$ in  Pb-Pb collisions on the value of the charm quark mass employed n the hadronization algorithm.}\label{Fig:ratio-vs-mc}
    \end{center}
\end{figure}
As anticipated, in the left panel of Fig.~\ref{Fig:FF} we show, for different collision centralities and transport coefficients, the fragmentation fractions into the different species of charmed hadrons obtained integrating the corresponding transverse momentum distributions. The charm fragmentation fractions in heavy-ion collisions provided by our model are quite insentitive to the centrality class and to the choice of transport coefficients. We also display, as a benchmark, the POWHEG-BOX predictions referring to $pp$ collisions at the same center-of-mass energy~\cite{Alioli:2010xd}. The most relevant finding is the enormous enhancement in the integrated yields of $D_s$ mesons and charmed baryons, in particular the heaviest ones, containing a strange diquark. In the POWHEG-BOX setup -- employing an old PYTHIA tune in the simulation of hadronization -- the production of these charmed hadrons is in fact suppressed by the need of exciting a $s\overline s$ or diquark-antidiquark pair from the vacuum, with a tunnelling probability suppressed by the high mass of these particles, which, on the contrary are quite abundant in the hot fireball produced in heavy-ion collisions. 

In Figs.~\ref{Fig:FF} (right panel) and~\ref{Fig:ratio-vs-mc} we display the sensitivity of our results to the value of the charm quark mass employed in the hadronization process (which, we remind, does not necessarily coincide with the one employed in the simulation of the initial hard $Q\overline Q$ production). One can see that, going from $m_c=1.5$ GeV (here our default choice) to $m_c=1.3$ GeV, one gets a larger (smaller) production of charmed mesons (baryons). This is essentially due to our resampling procedure. Taking $m_c\!=\!1.3$ GeV, with the values of diquark masses given in Table~\ref{tab:masses}, a non negligible fraction of charmed clusters carrying baryon number $B\!=\!1$ does not have a large enough invariant mass to allow their decay; hence one performs a resampling of the thermal particle with which the charm quark is recombined and this leads to a suppression of the charmed-baryon yields.

\begin{figure}[!h]
      \begin{center}
\includegraphics[clip,width=0.45\textwidth]{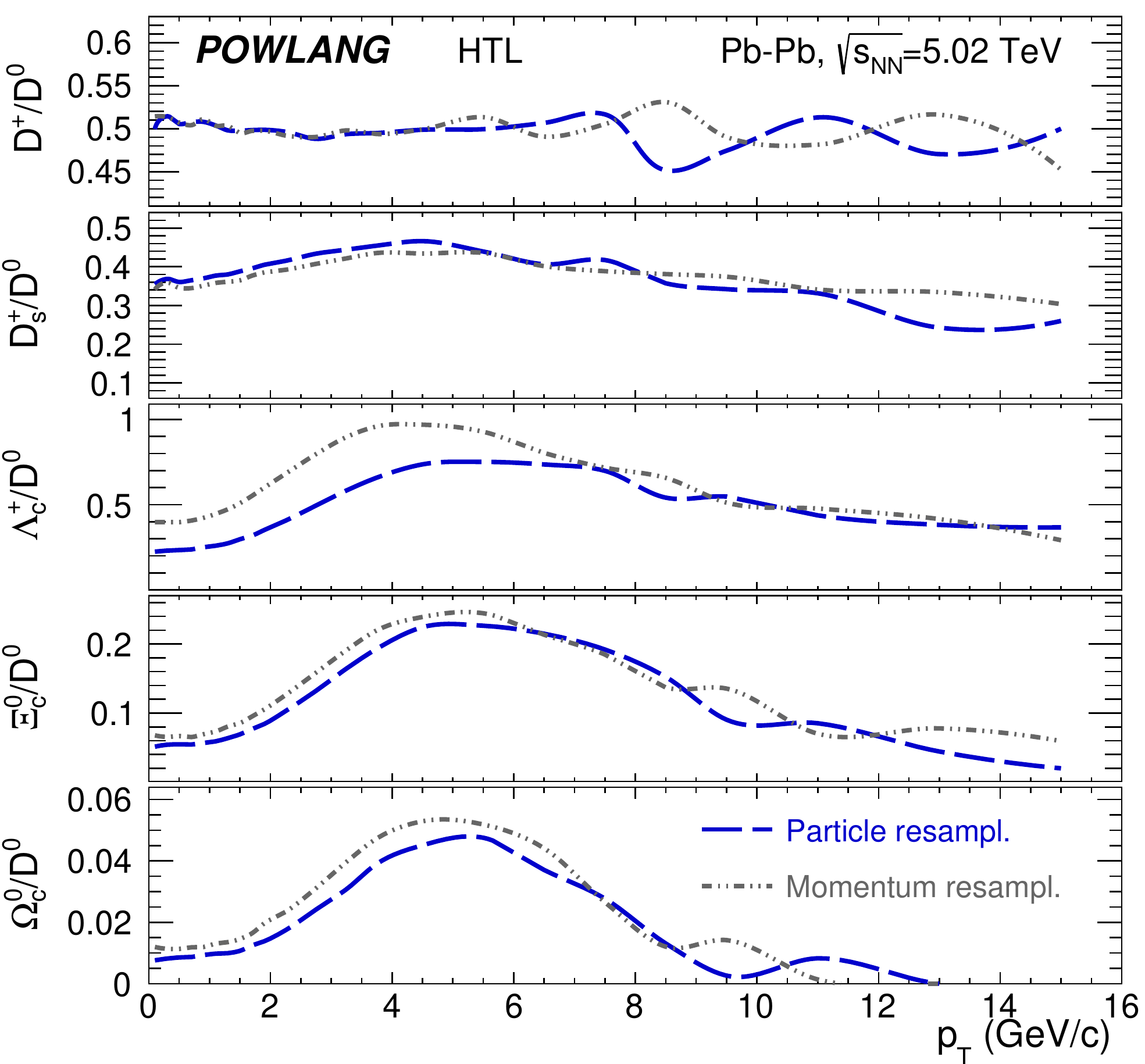}
\caption{Charmed hadron ratios as a function of $p_T$ in central Pb-Pb collisions for different resampling procedures of the light thermal particle involved in the recombination process.}\label{Fig:resampling}
    \end{center}
\end{figure}
We now wish to better quantify the systematic uncertainty introduced by the resampling procedure necessary when the cluster formed via recombination has not enough mass to undergo a two-body decay. Hence in Fig.~\ref{Fig:resampling} we compare the results obtained with our standard implementation -- corresponding to extracting a new thermal particle -- with the ones arising from a resampling of the momentum of the latter, keeping its species fixed and set by its thermal abundance. As one can see, in this second case one would get a larger production of charmed baryons.

\begin{figure*}[!h]
  \begin{center}
 \includegraphics[clip,width=0.45\textwidth]{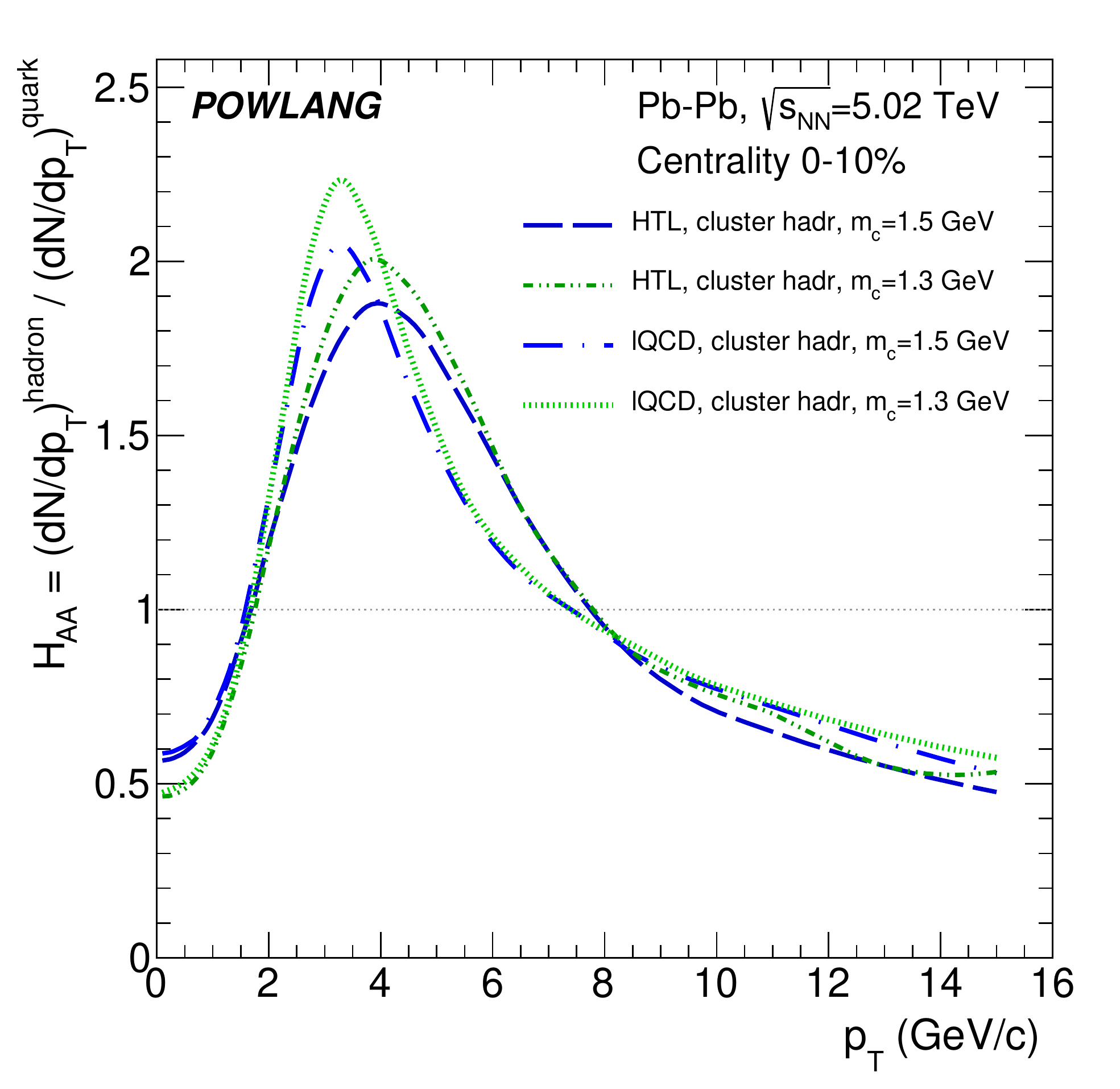}    
  \includegraphics[clip,width=0.45\textwidth]{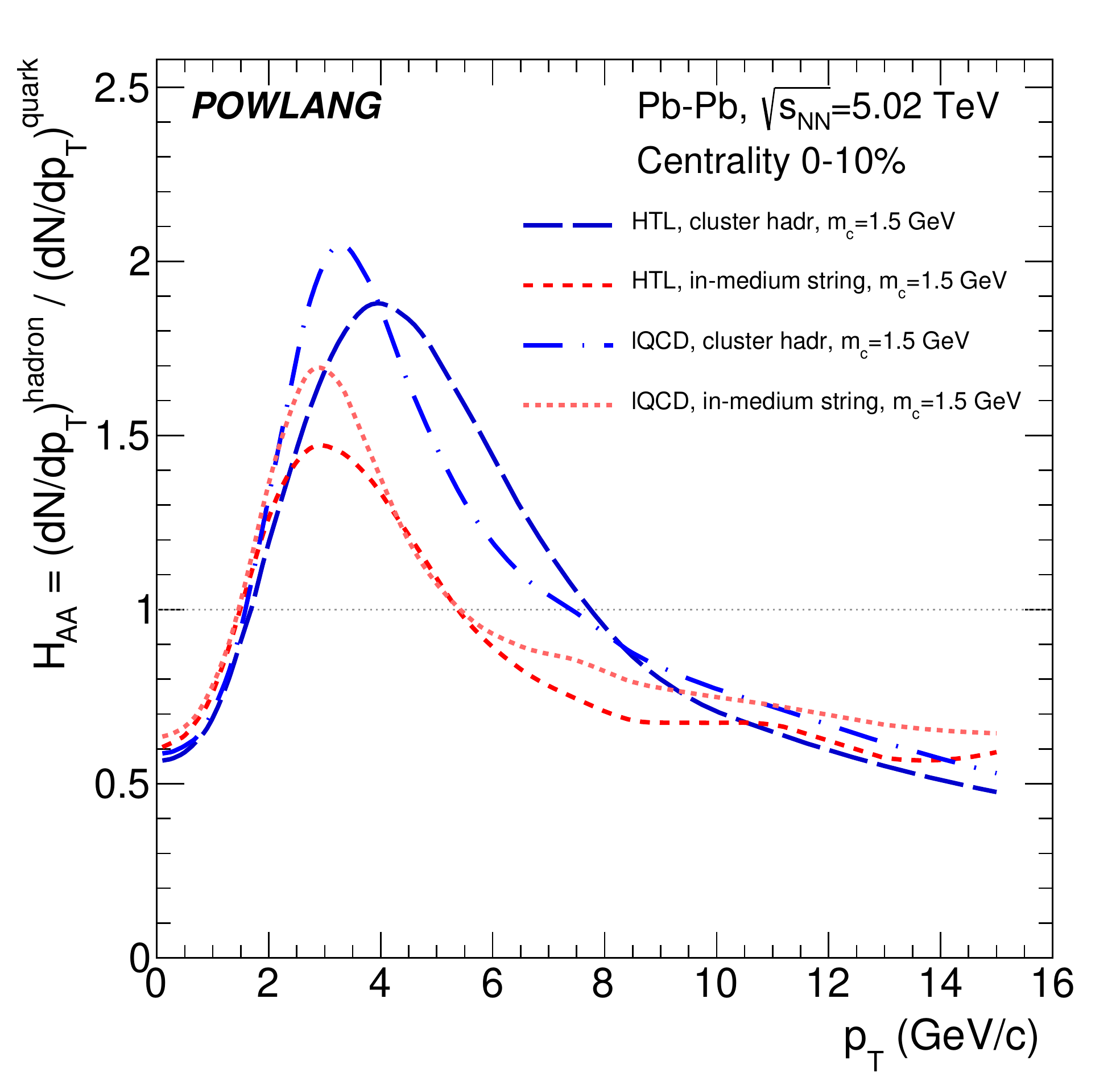}
\caption{Hadron-to-quark ratio of the charm $p_T$-spectra in central Pb-Pb collisions for different values of the charm quark mass (left panel) and different hadronization schemes (right panel) and two different choices of transport coefficients.}\label{Fig:HAA}
      \end{center}
      \end{figure*}
\begin{figure*}[!h]
  \begin{center} 
\includegraphics[clip,width=0.45\textwidth]{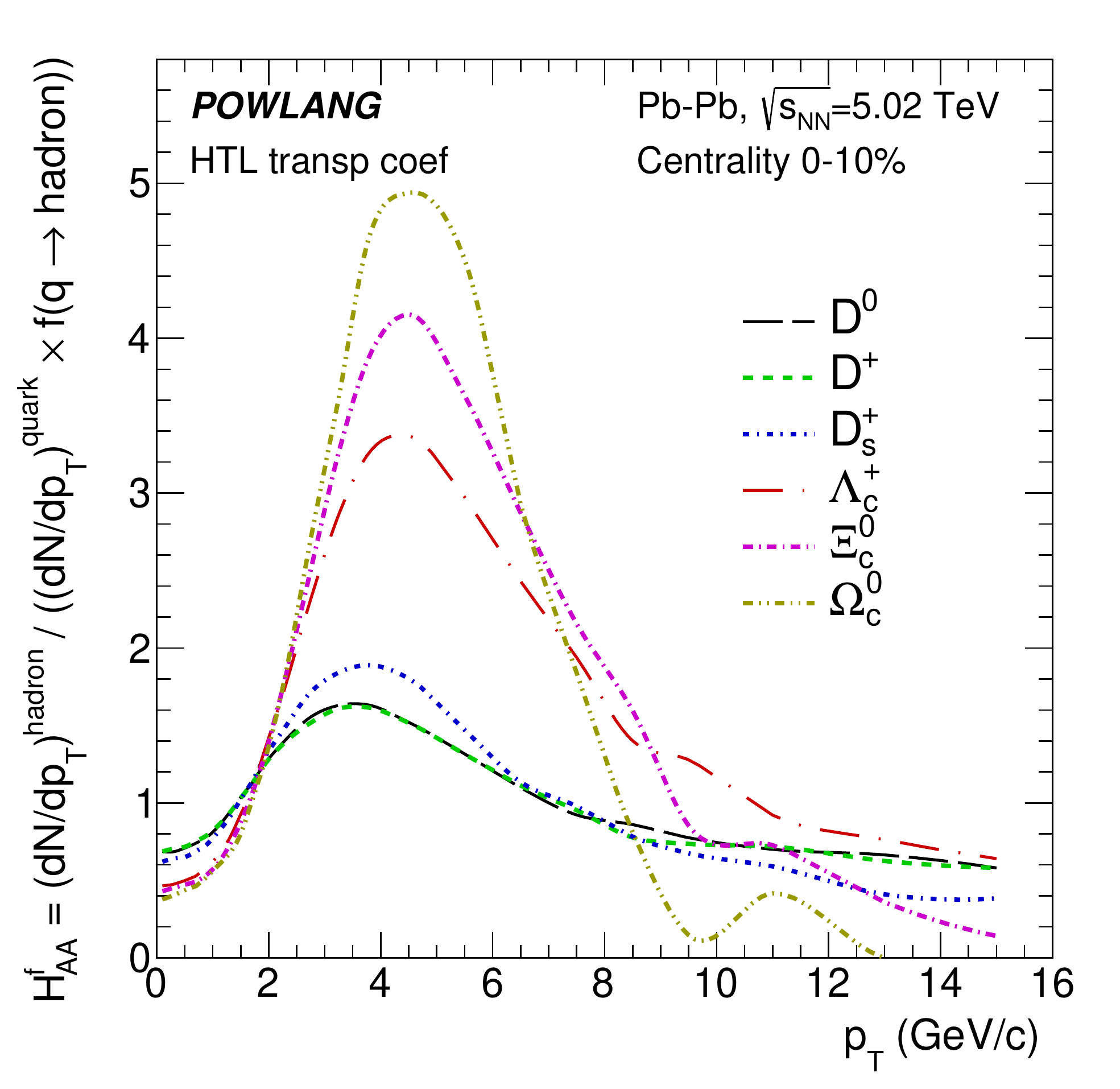}
\includegraphics[clip,width=0.45\textwidth]{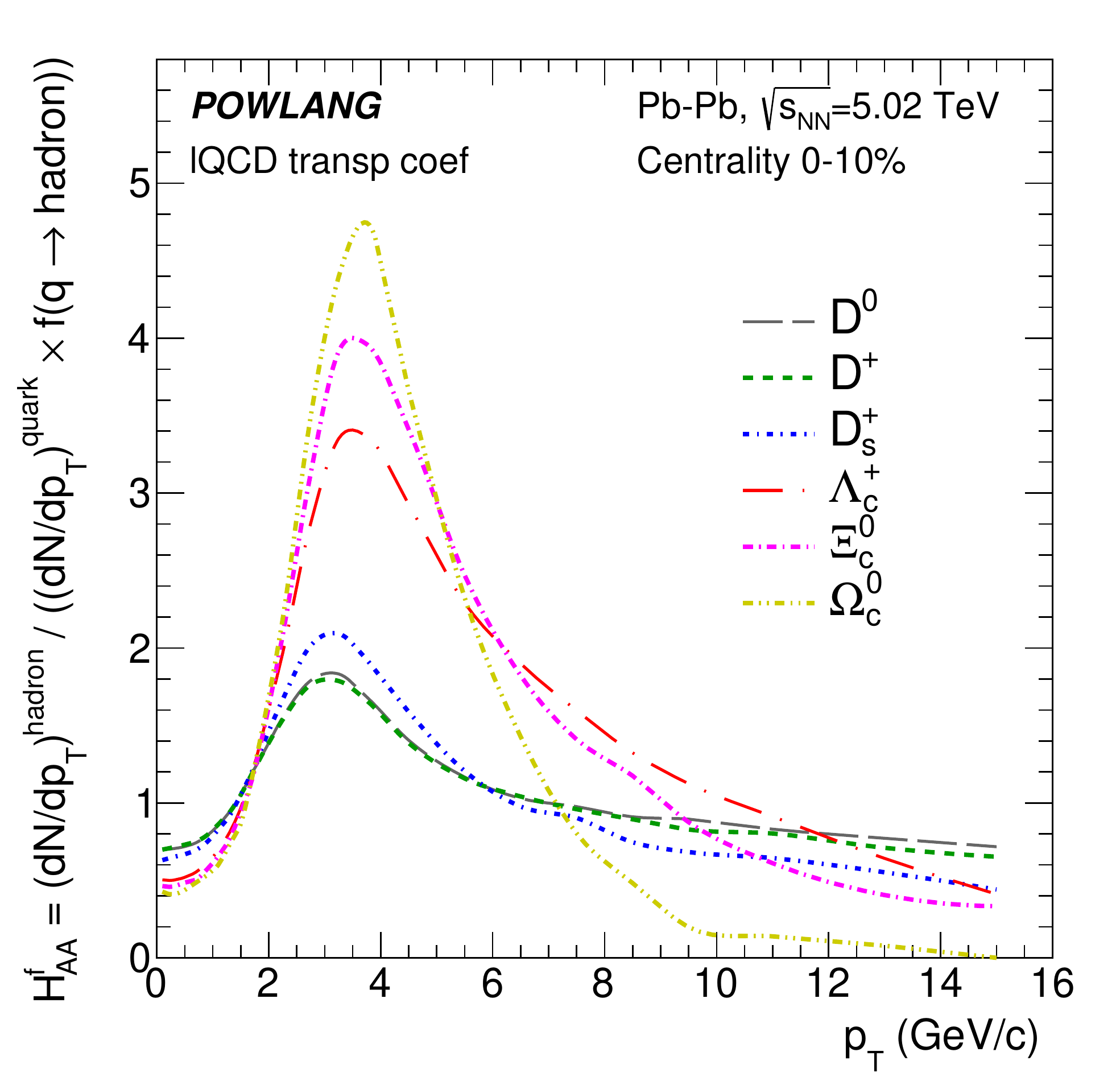}
\caption{Hadron-to-quark ratio of the charm $p_T$-spectra for various particle species in central Pb-Pb collisions for HTL (left panel) and lattice-QCD (right panel) transport coefficients. In order to compare spectra with the same normalization, the $p_T$-distribution of each hadronic species has been divided by its corresponding fragmentation fraction.}\label{Fig:HAA-species}
     \end{center}
\end{figure*}
We now study how in our model hadronization -- occurring via recombination with a thermal parton followed in most cases by a two-body decay -- affects the final momentum distributions of charmed hadrons. For this purpose, we introduce the following quantities, namely the ratio of the hadron and quark $p_T$ distributions
\beq
H_{\rm AA}\equiv\frac{(dN/dp_T)^{\rm all\,hadrons}}{(dN/dp_T)^{\rm quark}}
\eeq
and the one reffering to the single charmed hadron species
\beq
H_{\rm AA}^f\equiv\frac{(dN/dp_T)^{{\rm hadron}_f}}{(dN/dp_T)^{\rm quark}}\frac{1}{f(c\to{\rm hadron}_f)}\,,
\eeq
where in the last equation the hadron spectrum has been normalized to the corresponding fragmentation fraction, in order to compare distributions with the same normalization. Although not experimentally accessible, the two ratios allow one to quantify the effect of hadronization on the particle kinematics.\\
In Fig.~\ref{Fig:HAA} we study how the effect of hadronization on the final $p_T$-distributions depends on the charm quark mass, on the transport coefficients and on the modelling of the cluster/string decay or fragmentation. All the curves display the same qualitative behavior, namely a depletion of the soft-momentum region, a sizable enhancement for intermediate momenta and, finally, a quenching of the high-momentum tail. This comes from the processes which dominate in the different kinematic regions. The reshuffling of the momenta from low to intermediate $p_T$ is a consequence of the recombination process occurring on the hadronization hypersurface, through which the final charmed hadrons receive a further radial boost arising from the collective flow of the fireball, beside the one acquired through multiple scatterings of the heavy quarks in the deconfined medium. On the other hand, high-$p_T$ hadrons mostly come from the decay of very massive clusters, which in our model are fragmented as Lund strings, giving rise to a large number of hadrons; hence, on average, the final charmed hadron carries only a fraction of the momentum of the parent charm quark, as in the case of in-vacuum fragmentation. This, convoluted with a steeply-falling spectrum, entails a quenching of the hadronic distribution, leading to $H_{\rm AA}<1$. 
From the left panel of Fig.~\ref{Fig:HAA} one can see that the effect is pretty insensitive to the previous dynamics in the the deconfined fireball -- differences arising from the two choices of transport coefficients being small -- and to the value of the charm mass employed at hadronization. On the other hand, in the intermediate $p_T$-region differences between the two hadronization schemes (cluster decay vs string fragmentation) are non-negligible. In particular, cluster decay gives rise to a stronger enhancement of the final hadronic distributions at intermediate $p_T$, which display a larger amount of radial flow. This is due to the larger number of events in which hadronization occurs through a simple two-body decay, in which the daughter charmed hadron carries a larger fraction of the momentum of the parent cluster, as compared to the case in which a string undergoes a $N-$body fragmentation.\\
In Fig.~\ref{Fig:HAA-species} we display the $H_{\rm AA}^f$ for the different hadronic species. One can appreciate that
the reshuffling of the kinematic distributions from small to moderate $p_T$ is particularly important for charmed baryons, which receive a stronger boost from the radial flow of the fireball due to the higher mass of the thermal particle (a diquark) involved in the recombination process. For all hadron species the radial-flow peak turns out to be broader and shifted to higher $p_T$ in the case of HTL transport coefficients. 

\begin{figure*}[!ht]
       \begin{center}
\includegraphics[clip,width=0.45\textwidth]{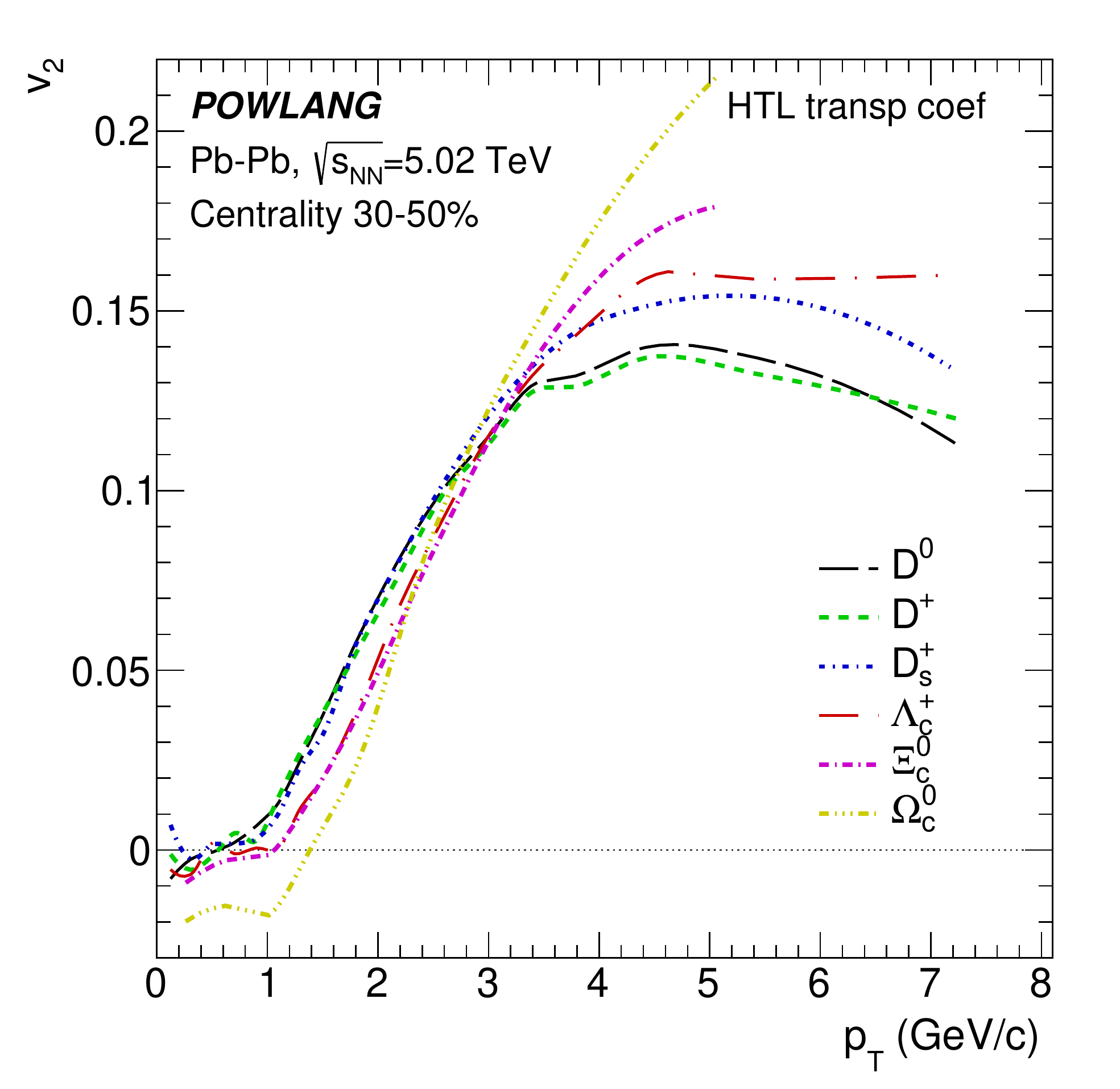}
\includegraphics[clip,width=0.45\textwidth]{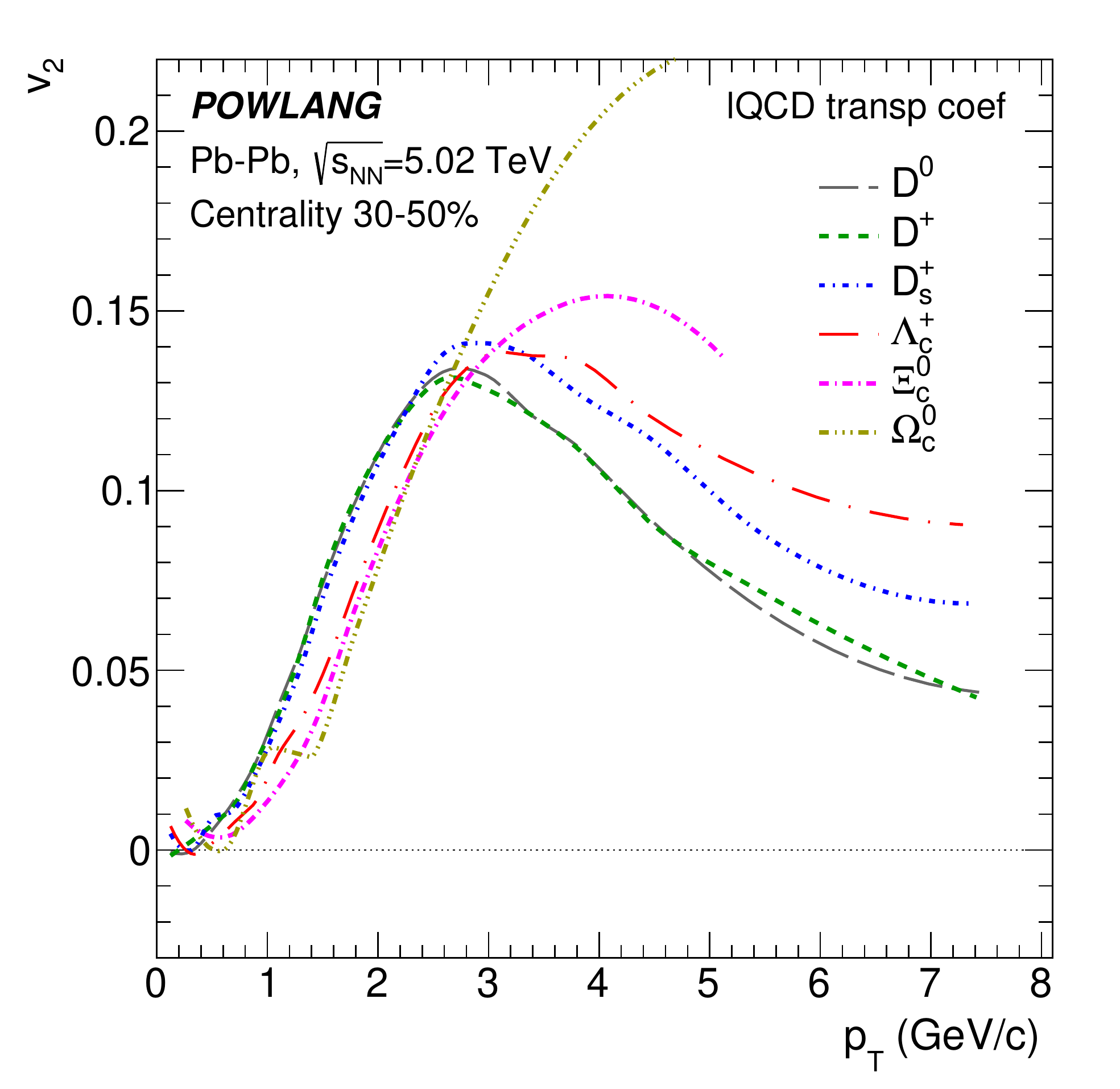}
\caption{The elliptic-flow coefficient $v_2$ of different charmed-hadron species in semi-central Pb-Pb collisions for HTL (left panel) and lattice-QCD (right panel) transport coefficients.}\label{Fig:v2_hadrons}
     \end{center}
\end{figure*}
\begin{figure}[!ht]
  \begin{center}
\includegraphics[clip,width=0.45\textwidth]{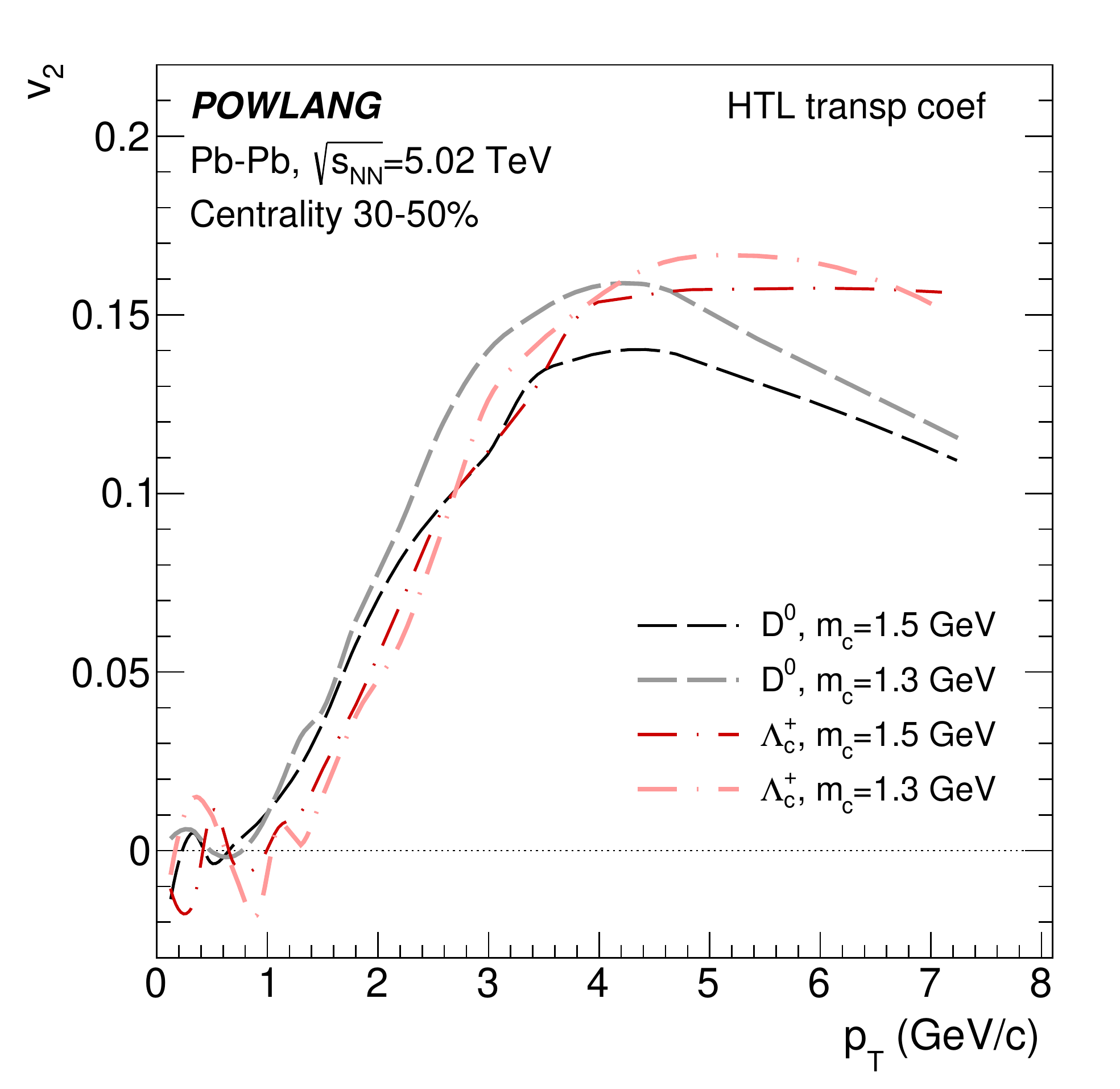}
\caption{The elliptic-flow coefficient $v_2$ of $D^0$ mesons and $\Lambda_c^+$ baryons in semi-central Pb-Pb collisions for different values of the charm mass.}\label{Fig:v2_mass}
\end{center}
\end{figure}
\begin{figure}[!ht]
  \begin{center}
\includegraphics[clip,width=0.45\textwidth]{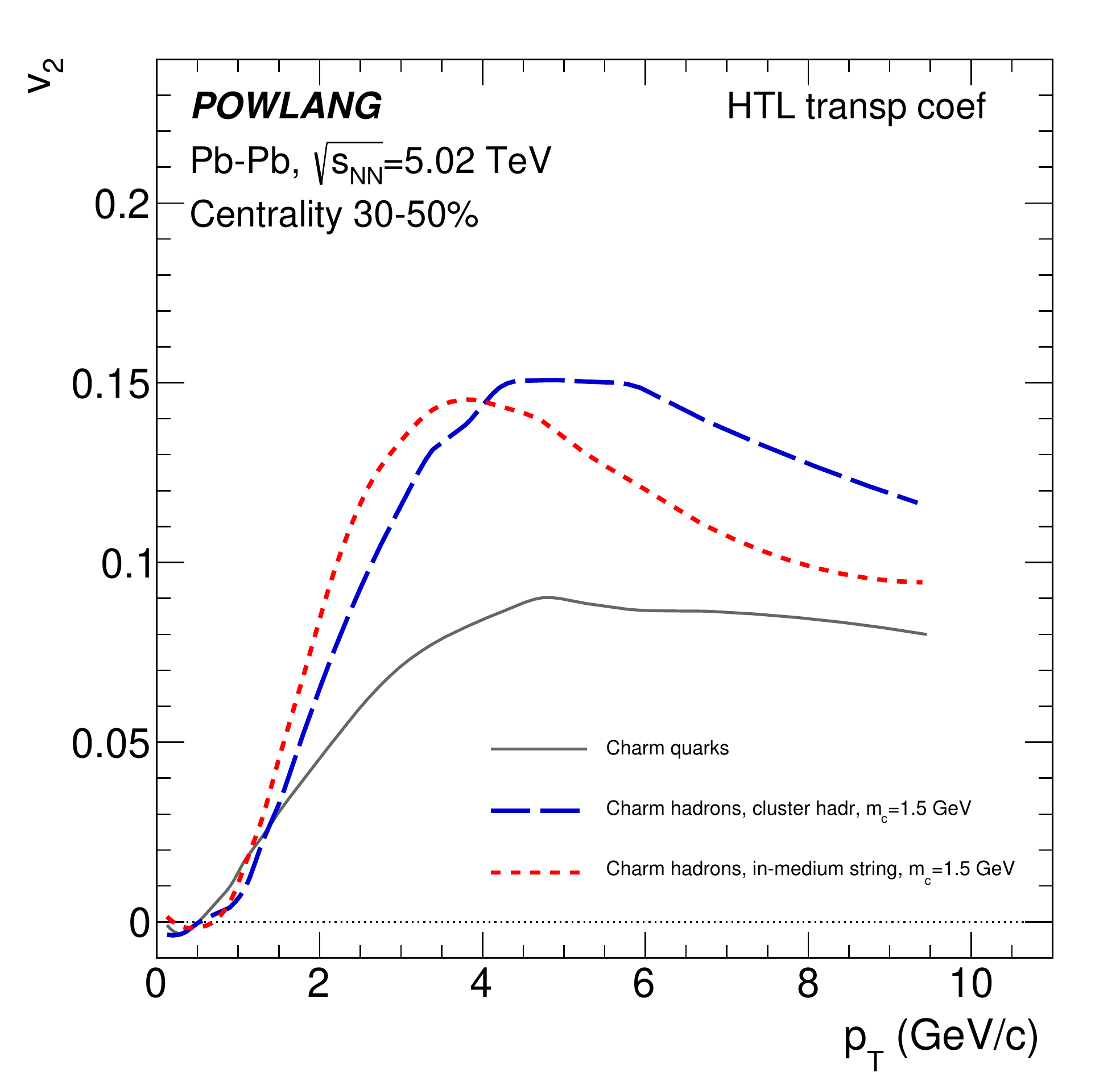}
    \caption{The elliptic-flow coefficient $v_2$ of charm quarks and hadrons in semi-central Pb-Pb collisions for two different in-medium hadronization schemes: cluster-decay vs string fragmentation.}\label{Fig:v2_q-vs-h}
\end{center}
\end{figure}
We now consider how our new hadronization scheme affects the azimuthal asymmetry of charmed particle emission. In Fig.~\ref{Fig:v2_hadrons} we display our results for the elliptic-flow coefficient $v_2$ in semi-central Pb-Pb collisions at $\sqrt{s_{\rm NN}}=5.02$ TeV for different charmed-hadron species. One can observe the mass ordering of the results, with the splitting of the meson and baryon curves which cross each others around $p_T\!\approx\!3$ GeV for both choices of transport coefficients. Within our model this splitting is only indirectly connected to the number of constituent quarks of the final hadrons. It rather arises from the different mass of the particle with which recombination takes place: an antiquark(quark) or a (anti)diquark for mesons and baryons, respectively.\\
In Fig.~\ref{Fig:v2_mass} we see that the meson-baryon splitting in the $v_2$ is present independently from the value of charm quark mass employed at hadronization.\\
In Fig.~\ref{Fig:v2_q-vs-h} we display the charm $v_2$ at the quark and hadron level, comparing for the hadrons the results obtained with the new cluster-decay approach to the ones based on the previous string-fragmentation algorithm. As one can see, about half of the final elliptic flow is acquired at hadronization. Results obtained in the two hadronization schemes are qualitatively similar, although in the cluster model the peak in the $v_2$ is shifted to slightly larger momenta. 

\begin{figure*}[!ht]
  \begin{center}
\includegraphics[clip,width=0.45\textwidth]{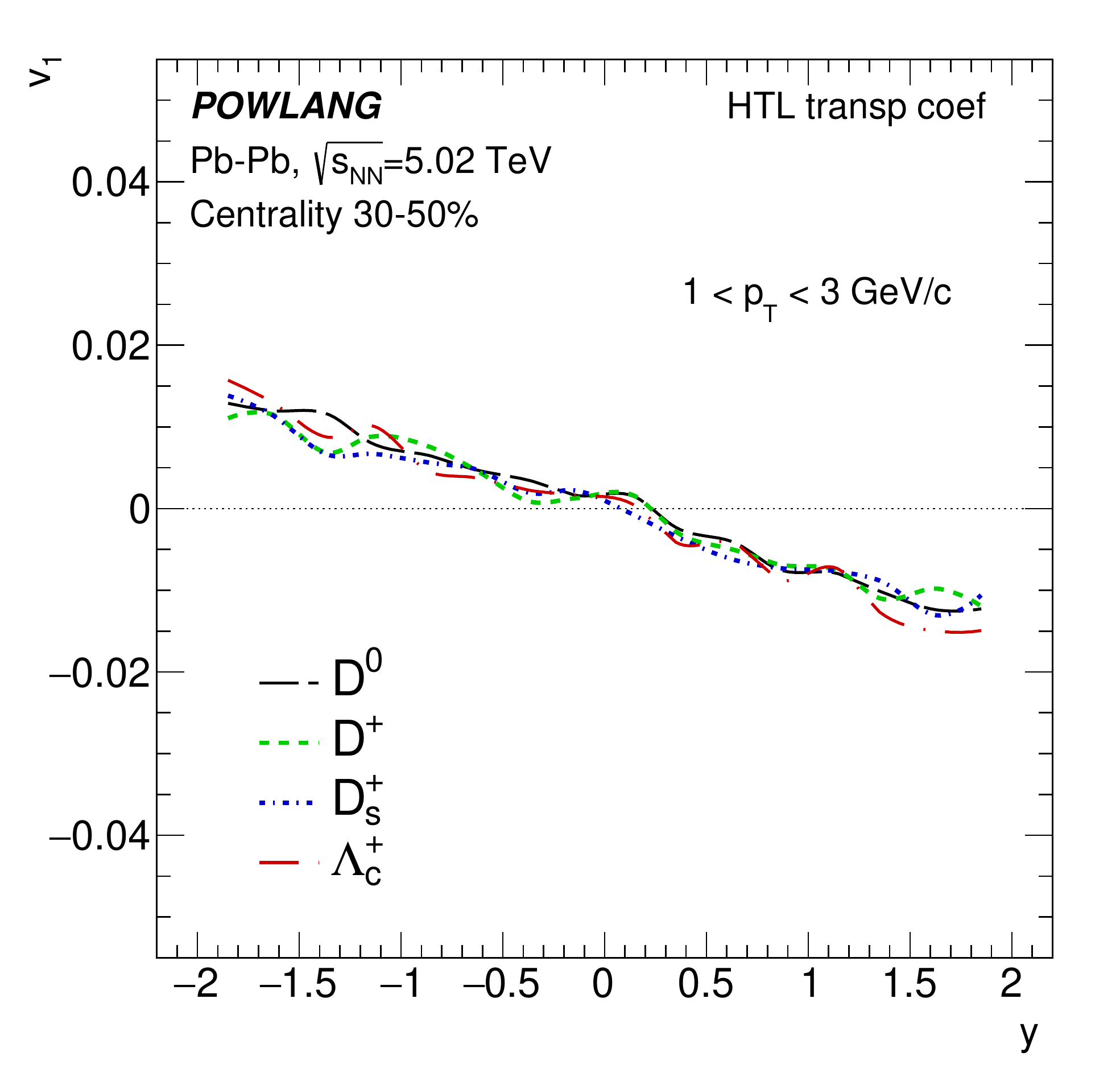}
\includegraphics[clip,width=0.45\textwidth]{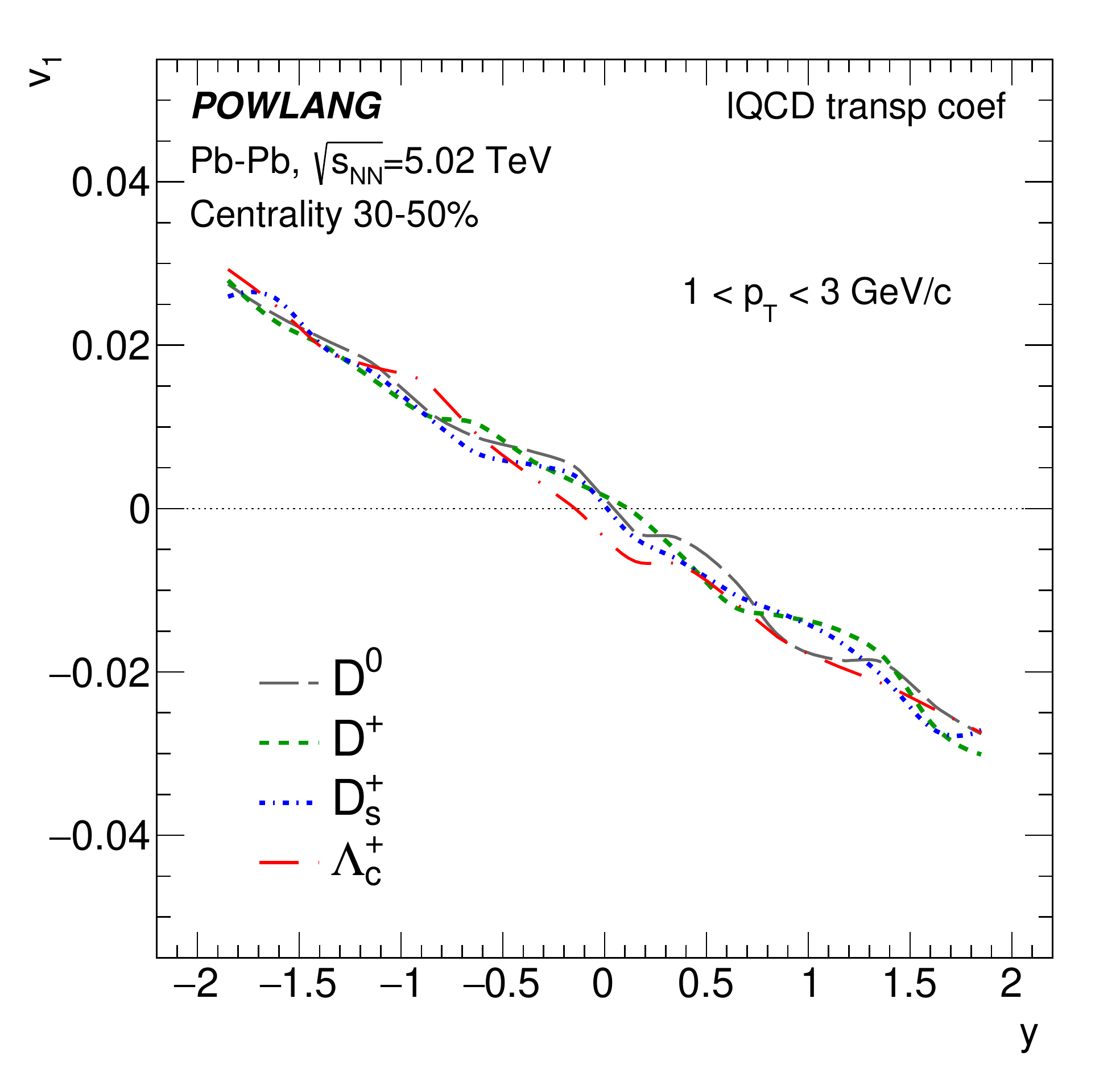}
\caption{The directed-flow coefficient $v_1$ of different charmed-hadron species as a function of rapidity in semi-central Pb-Pb collisions for HTL (left panel) and lattice-QCD (right panel) transport coefficients.}\label{Fig:v1_hadrons}
\end{center}
\end{figure*}
\begin{figure}[!ht]
  \begin{center}
\includegraphics[clip,width=0.45\textwidth]{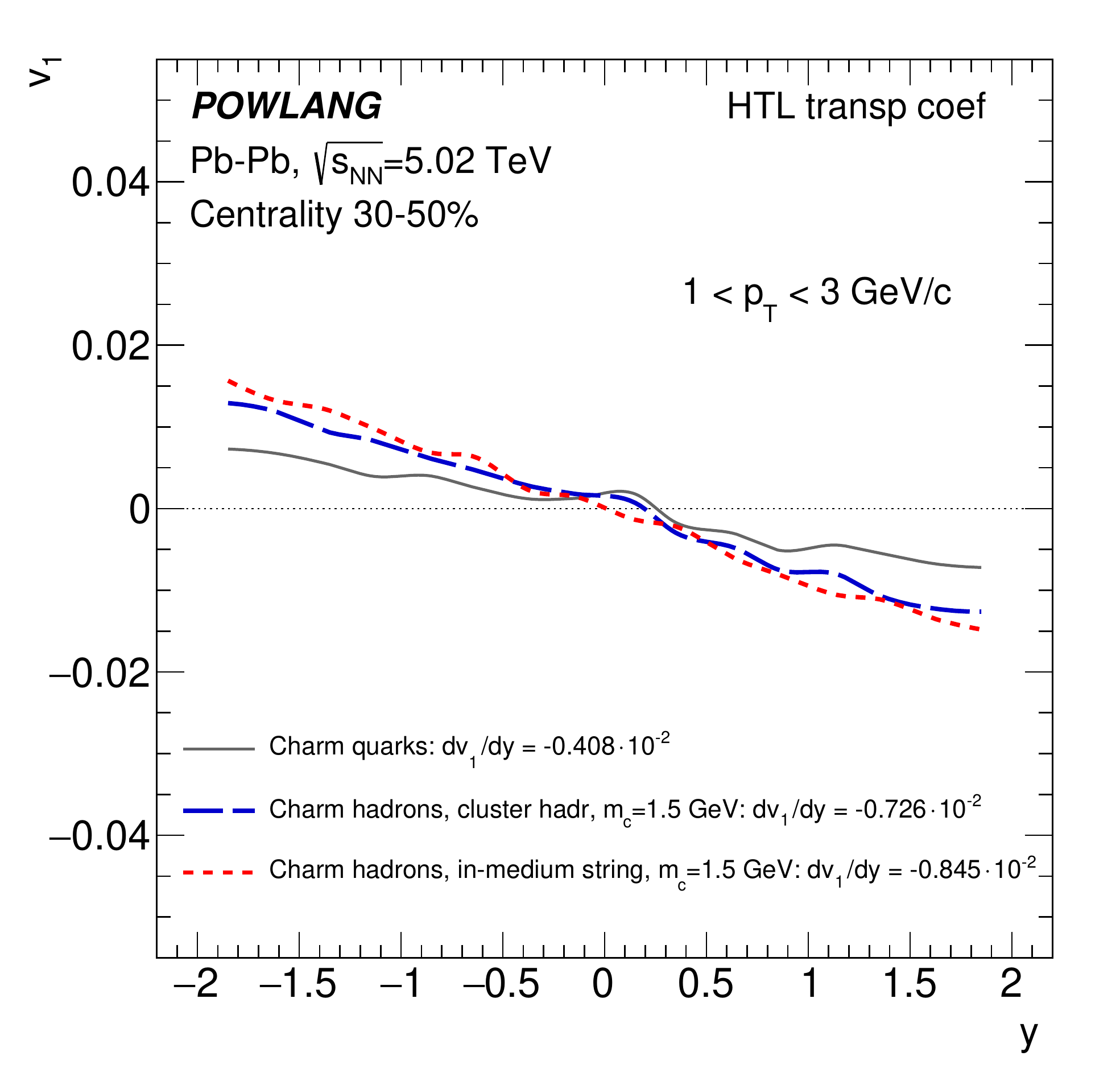}
\caption{The directed-flow coefficient $v_1$ of charm quarks and hadrons as a function of rapidity in semi-central Pb-Pb collisions for two different in-medium hadronization schemes: cluster-decay vs string fragmentation.}\label{Fig:v1_q-vs-h}
\end{center}
\end{figure}
Finally, in Figs.~\ref{Fig:v1_hadrons} and~\ref{Fig:v1_q-vs-h} we show our results for the directed-flow coefficient $v_1$ of charmed hadrons as a function of rapidity in semi-central Pb-Pb collisions. The origin of the charm directed flow, which turns out to be much larger than the one of light hadrons, has been discussed in detail in Refs.~\cite{Chatterjee:2017ahy,Oliva:2020doe,Beraudo:2021ont}. In particular, in Fig.~\ref{Fig:v1_q-vs-h} in which the slopes $dv_1/dy|_{y=0}$ of the quark and hadron curves are quoted, the effect of hadronization which increases the final signal can be appreciated. The fact that a significant contribution from hadronization is present also in a situation in which the $v_1$ of the light thermal partons of the background medium is very small is due to the strong space-momentum correlation entailed by construction by our local recombination process and has been discussed in detail in our past publication~\cite{Beraudo:2021ont}.

\begin{figure*}[!ht]
  \begin{center}
    \includegraphics[clip,width=0.95\textwidth]{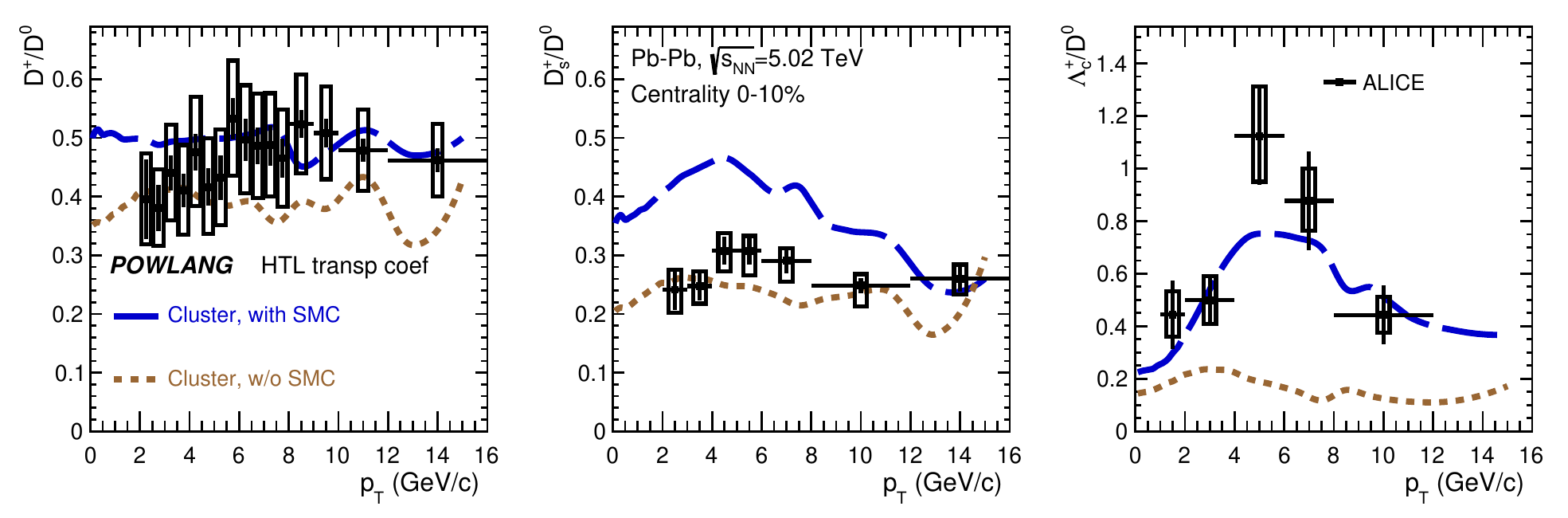}
\caption{Charmed hadron ratios with (dashed blue curves) and without (dotted brown curves) space-momentum correlations. The last case is obtained mixing momentum and position of heavy quarks at hadronization from different events.}\label{Fig:yields-noSMC}
\end{center}
  \end{figure*}
\begin{figure}[!ht]
  \begin{center}
\includegraphics[clip,width=0.45\textwidth]{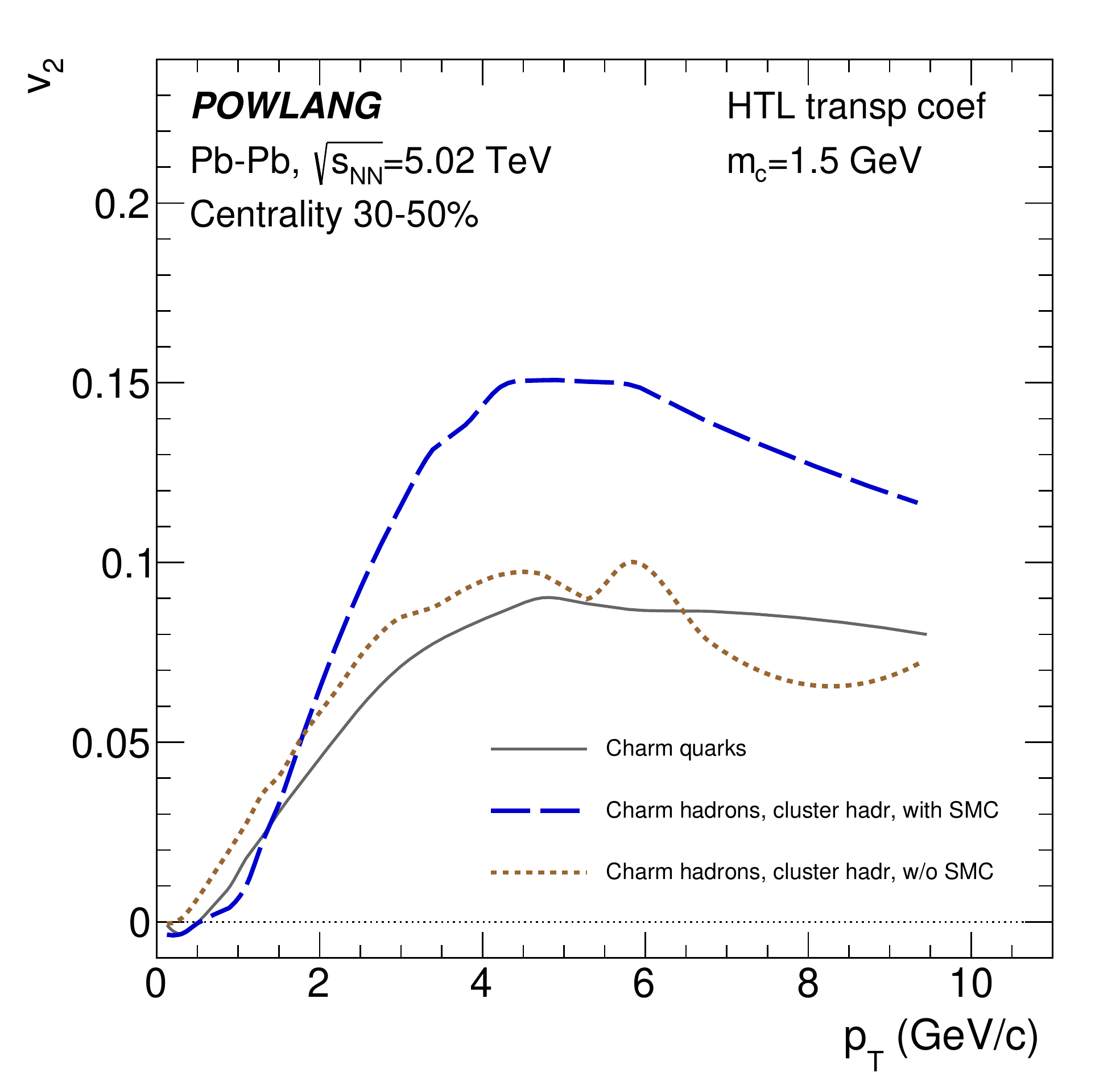}
\caption{Charmed hadron elliptic-flow coefficient $v_2$ with (dashed blue curve) and without (dotted brown curve) space-momentum correlations compared to the one of charm quarks (continuous black curve).}\label{Fig:v2-noSMC}
\end{center}
\end{figure}
After displaying the predictions obtained interfacing this newly developed hadronization model to our transport setup, before summarizing our major findings we wish to devote a special attention to the above mentioned issue of space-momentum correlations (SMC's). As already discussed, the formation of a color-singlet cluster/string in our model is a process which occurs \emph{locally}, with probability 1, via recombination of the charm quark with a light thermal particle from the same fluid cell. Since the direction of motion of the heavy quark is correlated with its position on the hadronization hypersurface and since the same holds for the collective velocity of the fluid-cell it occupies, it follows that the momenta of the heavy quark and of the light particle involved in the recombination process tend to be quite collinear in the laboratory frame. First of all, as already shown in Sec.~\ref{sec:model}, this affects the invariant mass of the produced color-singlet cluster, in most of the cases quite small, and accordingly the hadrochemistry of its decay products. Secondly, both the radial flow and the azimuthal asymmetry coefficients $v_n$ of the daughter hadrons get enhanced with respect to the partonic ones. Here we try to better quantify such an effect. For this purpose we compare the results for the yields and kinematic distribution of the final charmed hadrons obtained with the standard implementation of our model, including by construction SMC's, to the ones in which SMC's are suppressed through the event-mixing technique described in Sec.~\ref{sec:model}. 

In Fig.~\ref{Fig:yields-noSMC} we compare the yields of charmed hadrons, plotted as a function of $p_T$ and normalized to those of $D^0$ mesons, in central Pb-Pb collisions with (dashed blue curves) and without (dotted brown curves) SMC's. As one can see, in the second case the production of $D_s^+$ mesons and $\Lambda_c^+$ baryons is slightly and strongly suppressed, respectively. Furthermore, the peak in the $\Lambda_c^+/D^0$ ratio at $p_T\approx 5$ GeV, attributed to the larger radial flow of charmed baryons, disappears: in fact, breaking SMC's, the charm quark is no longer recombined with a thermal particle moving on average in the same direction.
The different final hadrochemistry obtained with and without SMC's arises instead from the different invariant mass of the parent clusters. As above discussed (see the right panel of Fig.~\ref{fig:Mdistr}), SMC's favor the formation of small-mass clusters undergoing a two-body decay in which the daughter charmed hadron carries the strangeness and baryon number of the parent. On the contrary, removing SMC's enhances the fraction of high-mass clusters which are fragmented as Lund strings. This leads to a suppression of charmed hadrons carrying non-zero strangeness and/or baryon number. In fact, their formation requires the recombination of a $c$ quark with a $\overline s$ antiquark or with a light diquark, either already there or excited from the vacuum. In the case of cluster decay this corresponds to a penalty factor arising simply from their relative thermal abundance $\sim e^{-m/T}$. On the other hand, in the case of string fragmentation the production requires the excitation from the vacuum of a $q\overline q$ or a diquark-antidiquark pair, with a tunnelling probability suppressed by a factor $e^{-m^2/\pi\kappa}$ ($\kappa$ being the string tension). As a result, in the absence of SMC's the $\Lambda_c^+/D^0$ ratio turns out to be almost independent of $p_T$ and close to the value measured in $e^+e^-$ collisions.  

It is also interesting to check how much SMC's contribute to the azimuthal asymmetry of the distributions of final-state particles. For this purpose, in Fig.~\ref{Fig:v2-noSMC} we display the elliptic flow coefficient $v_2$ of charmed hadrons in semi-central Pb-Pb collisions corresponding to the default implementation of our hadronization model including SMC's (dashed blue curves), compared to the results obtained in their absence (dotted brown curves) and to the charm quark $v_2$. As one can see, SMC's are responsible for a strong enhancement of the elliptic flow of final charmed hadrons, while breaking these correlations leads to a result very close to the one obtained at the quark level.

\section{Discussion and perspectives}\label{sec:discussion}
\begin{figure}[!ht]
  \begin{center}
    \includegraphics[clip,width=0.45\textwidth]{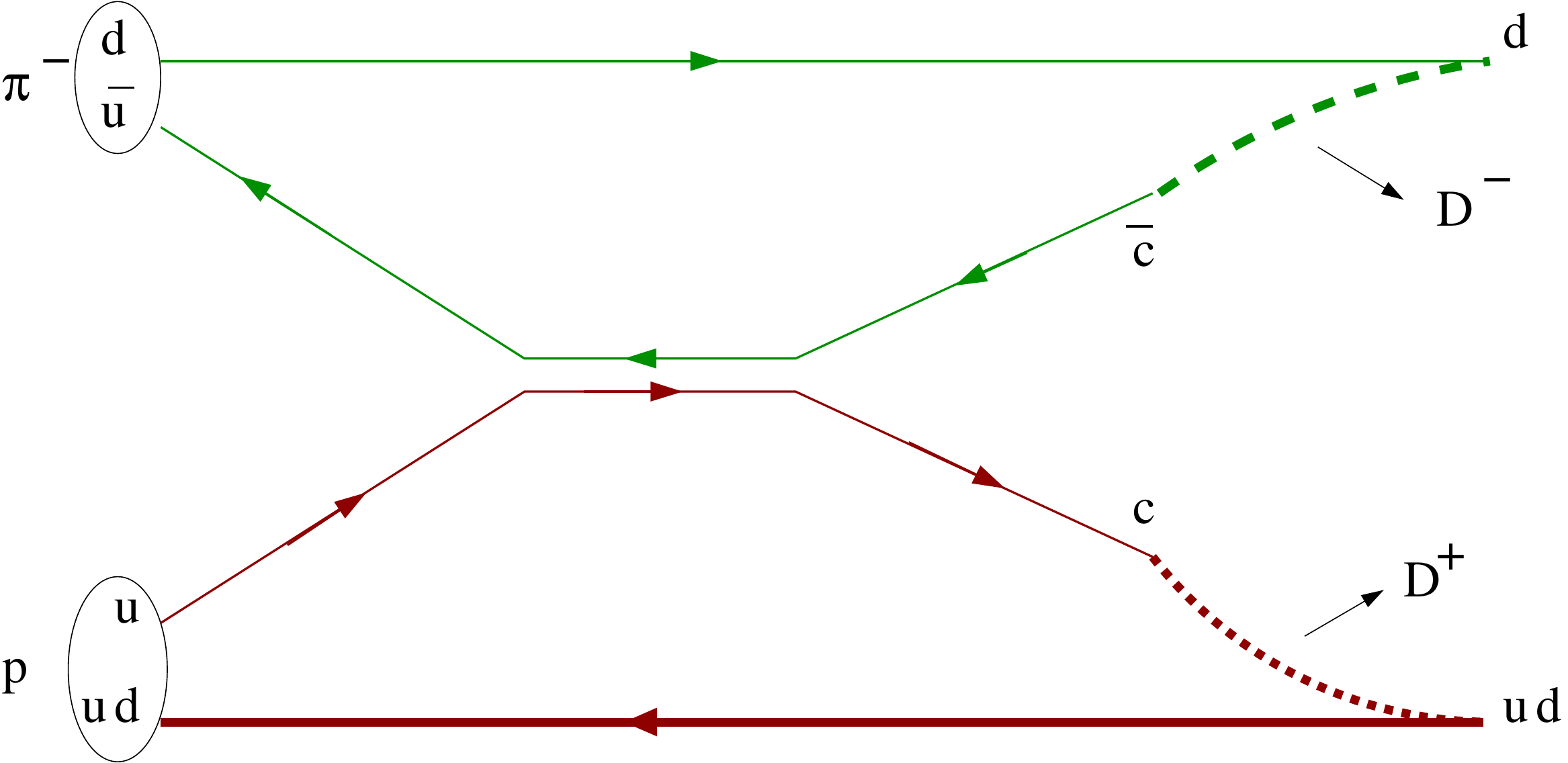}
    \caption{Recombination of charm quarks with the beam-remnant explaining the $D^-/D+$ asymmetry in $\pi^-p$ collisions.}\label{Fig:beam-drag}
        \end{center}
\end{figure}

The major purpose of our work was to develop a novel hadronization algorithm to interface to heavy-quark transport simulations capable of reproducing recent non-trivial experimental findings like, for instance, the enhancement of the baryon-to-meson observed (also) in the charm sector in high-energy nucleus-nucleus and proton-proton collisions.
We developed a picture in which a heavy quark, after reaching the hadronization hypersurface at the end of its propagation in the deconfined fireball, gives rise to a color-singlet cluster formed via recombination with a light antiquark or diquark from the background hot medium. The cluster undergoes then a (two-body, if sufficiently light) hadronic decay. The process is assumed to occur locally, within the same fluid cell, whose collective velocity makes the momentum of the light thermal particle aligned with the one of the charm quark. This space-momentum correlation turns out to be fundamental to transfer the collective flow of the fireball to the final charmed hadrons and to suppress the formation of high-mass clusters, thus impacting on the relative abundances of the different charm hadron species arising from their decays. As discussed in the text, this is particularly relevant in order to obtain charm-quark fragmentation fractions in nuclear collisions in agreement with the most recent experimental data.  

The details and the few free parameters of our model have been discussed at length in the text. Here we prefer to emphasize some very general aspects of the latter, necessary to reproduce otherwise puzzling experimental observations, like the enhanced production of charmed baryons in high-energy nuclear (and proton-proton) collisions and the strong signatures of collective flow found in the kinematic distributions of charmed hadrons. We wish to stress that, actually, some of these general features are somehow shared by other apparently very different descriptions of hadronization, like coalescence calculations or string models and color-reconnection algorithms implemented in multi-purpose event generators like PYTHIA.

First of all the recombination occurs with some low-$p_T$ parton which is already present in the medium, close in space to the hadronizing heavy quark, giving rise to the formation of a quite small invariant-mass object. In coalescence approaches this is imposed by the width of the hadron Wigner function, which favors the recombination of partons which are close in space and having collinear momenta. In QCD event generators this occurs, for instance, when color neutralization takes place stretching a string connecting partons produced in the hard event with a beam-remnant. Such a string tends to boost the hadrons produced in its fragmentation towards the direction of the beam, in analogy with the transverse-momentum reshuffling arising from the collective flow of the fireball in heavy-ion collisions due to recombination. As already mentioned in the Introduction, this beam-remnant drag can give rise to rapidity-dependent particle/antiparticle asymmetries, which have been experimentally observed~\cite{WA82:1993ghz,E769:1993hmj,E791:1996htn,WA89:1998wdl,E791:1997eip,SELEX:2001iqh,D0:2015rnb}. An example is illustrated in Fig.~\ref{Fig:beam-drag}, where we display the color-flow of a $c\overline c$-production event in a $\pi^-p$ collision. These processes were studied in the past in several fixed-target experiments at CERN and at Fermilab~\cite{WA82:1993ghz,E769:1993hmj,E791:1996htn,WA89:1998wdl}, finding a $D^-/D^+>1$ ratio at forward rapidity, along the direction of the pion beam. These data were successfully interpreted within the string hadronization model~\cite{Norrbin:1998bw,Norrbin:2000zc}: at very forward rapidity the $\overline c$ antiquark produced in the hard event and the $d$ quark from the pion remnant are quite collinear and lead to the production of a small invariant-mass cluster which can only undergo a collapse into a single $D^-$ meson, conserving the flavor of the original partons. Hence, the formation of small-mass color-singlet objects via parton recombination is the key to understand the modifications of relative hadron-yields when studying particle production in very different colliding systems, the only difference being the reservoir from which the second parton is taken from: a hot deconfined medium and the beam remnant in heavy-ion and elementary hadronic collisions, respectively.  

A crucial assumption in our hadronization model is that -- in analogy with some coalescence study~\cite{Oh:2009zj} -- a heavy quark can undergo recombination not only with a light antiquark, but also with a diquark, assumed to be present in the medium around the QCD pseudocritical temperature. This is crucial in order to reproduce the enhanced charmed baryon production observed in the data. The recombination with diquarks gives rise also to a characteristic splitting of the $v_2$ curves of charmed baryons with respect to the ones of $D$ mesons, which has to be attributed to the larger mass of the diquarks.  

As mentioned in the Introduction, some recent measurements performed in $pp$ collisions showed features which are usually attributed, in a heavy-ion environment, to the formation of a hot strongly-interacting medium, like signatures of collective flow or modification of the hadrochemistry (i.e. strange particle production and baryon/meson ratio). Attempts were performed to reproduce these observations by implementing in QCD event generators some amount of color reconnections~\cite{OrtizVelasquez:2013ofg,Christiansen:2015yqa,Bierlich:2015rha}: the idea is that partons are no longer connected by strings according to their color charge in the final stage of the simulation, but in order to minimize the invariant mass of the strings.
Hence, our implementation of parton recombination --  assumed to occur locally and giving rise, due to SMC's, to small-mass clusters -- can be seen as an extreme case of color reconnection. 

So far, our new model of hadronization has been applied only to the study of heavy flavor production in relativistic heavy-ion collisions, since in this case there is quite a strong evidence that a hot deconfined medium is actually formed.
Also in our past works medium effects on heavy-quark propagation and hadronization were introduced only in nucleus-nucleus and proton-nucleus collisions. The proton-proton case was considered just as a benchmark, with the $Q\overline Q$ hard production, parton-shower and hadronization stage simulated through the POWHEG-BOX package, assuming no medium is formed in these collisions. 
However, as already discussed, non-trivial effects like an enhanced relative production of charmed baryons have been recently observed also in proton-proton collisions~\cite{ALICE:2020wfu,ALICE:2020wla,ALICE:2021rzj}.
Hence it will be interesting to extend our study also to the $pp$ case, accounting for the possibility that, also in this case, a small QGP droplet is formed.
In light of the small dimension and short lifetime of the system we do not expect major modifications to the heavy-quark kinematic distributions to be introduced by the medium. However, the nearby deconfined light partons can surely play a role at hadronization, as suggested by the measured charmed hadron yields.
Clearly, this requires having at our disposal a validated model of the initial conditions for the hydrodynamic evolution of the background medium, both for the case of minimum-bias and high-multiplicity $pp$ collisions. Various options can be explored, all more or less based on the idea of eikonal entropy deposition by participant/colliding constituent quarks of via a ``reduced-thickness'' function~\cite{Bozek:2016kpf,Mitchell:2016jio,dEnterria:2010xip,Moreland:2014oya,Moreland:2018gsh}.
Investigations are currently in progress and we postpone this more complete study, which would give us access also to other observables like the nuclear modification factor of the momentum distributions of charmed hadrons, to a future publication.

\appendix
\section{Medium modelling}\label{App:hydro}
We assume that the heavy quarks produced in high-energy nuclear collisions propagate through a deconfined medium undergoing a (3+1)D hydrodynamic evolution. To model the initial condition of the latter, following Ref.~\cite{Bozek:2010bi}, we employ the optical Glauber model, assuming that the initial entropy deposition arises from a linear combination of the contributions from the participant (``wounded'') nucleons and from the binary nucleon-nucleon collisions, with relative weights $(1-\alpha_h)$ and $\alpha_h$, respectively. We assume that at the initial longitudinal proper time $\tau_0$ the fireball has a finite extension in rapidity ($\eta_s\equiv\frac{1}{2}\ln\frac{t+z}{t-z}$) modelled by the function
\beq
  H(\eta_s)=\exp\left[-\dfrac{(|\eta_s|-{\eta_{\rm flat}})^2}{2{\sigma_{\eta}}^2}\theta(|\eta_s|-{\eta_{\rm flat}})\right],\label{Eq:flat}
\eeq
with a central flat plateau of extension 2$\eta_{\rm flat}$ beyond which the density drops to zero with a Gaussian smearing $\sigma_\eta$. Furthermore right (left)-moving wounded nucleons are assumed to produce more particles at forward (backward) rapidity, respectively. The effect is parametrized by the function
\beq
f_{+/-}(\eta_s)=
\begin{cases}
0/2 & \eta_s < -\eta_m\\
\dfrac{\pm\eta_s+{\eta_m}}{{\eta_m}} & -\eta_m \le \eta_s \le \eta_m\\
2/0 & \eta_s > \eta_m.\label{Eq:tilt}
\end{cases}
\eeq
Notice that this is responsible for the tilting of the fireball and for the developing of a non-vanishing directed flow quantified by the $v_1$ coefficient.
The initial entropy density at the hydrodynamization time $\tau_0$ is then given by\\
\begin{strip}
  \begin{equation}
s(\vec x_\perp,\eta_s;b)=s_0\frac{(1\!-\!\alpha_h)[n_{\rm part}^A(\vec x_\perp;b)f_+(\eta_s)\!+\!n_{\rm part}^B(\vec x_\perp;b)f_-(\eta_s)]+\alpha_h n_{\rm coll}(\vec x_\perp;b)}{(1\!-\!\alpha_h)n_{\rm part}(\vec 0_\perp;0)+\alpha_h n_{\rm coll}(\vec 0_\perp;0)}H(\eta_s).\label{Eq:entropy}
    \end{equation}
\end{strip}
The values of the parameter entering into Eqs.~(\ref{Eq:flat})-(\ref{Eq:entropy}) and allowing a satisfactory description of soft observables can be found in our previous publication~\cite{Beraudo:2021ont}. After initialization, the hydrodynamic equations describing the medium evolution are solved through the ECHO-QGP code~\cite{DelZanna:2013eua}.

\section{Heavy-quark production and transport}\label{App:transport}
The initial $Q\overline Q$ production during the crossing of the two nuclei is generated through the POWHEG-BOX package~\cite{Frixione:2007nw,Alioli:2010xd}, taking care both of the hard process and of the initial and final-state parton shower stage. EPS09 nuclear parton distribution functions~\cite{Eskola:2009uj} are employed in the simulations.
Heavy quarks are then distributed in the transverse plane according to the local density of binary nucleon-nucleon collisions. As discussed in detail in previous publications their subsequent transport in the QGP is described through {the following relativistic Langevin equation~\cite{Beraudo:2014boa,Beraudo:2017gxw}
\beq
{\Delta \vec{p}}/{\Delta t}=-{\eta_D(p)\vec{p}}+{\vec\xi(t)}\label{eq:Langevin}
\eeq
containing a deterministic friction force quantified by the drag coefficient $\eta_D$ and a random noise term specified by its temporal correlator
\beq
\langle\xi^i(\vec p_t)\xi^j(\vec p_{t'})\rangle\!=\!{b^{ij}(\vec p_t)}{\delta_{tt'}}/{\Delta t}\;,
\eeq
with
\beq
{b^{ij}(\vec p)}\!\equiv\!{\kappa_\|(p)}\hat p^i\hat p^j+{\kappa_\perp(p)}(\delta^{ij}\!-\!\hat p^i\hat p^j).
\eeq
Different values for the longitudinal/transverse momentum-diffusion coefficients $\kappa_\|$ and $\kappa_\perp$ are employed}, arising either from a weak-coupling calculation with resummation of medium effects (HTL scheme) or from lattice-QCD simulations (l-QCD scheme). Both schemes display some shortcomings due to the mismatch between the kinematic and temperature conditions of experimental relevance and the ones in which theoretical calculations can provide results based on solid grounds. {The drag coefficient $\eta_D$ is connected to the momentum-diffusion coefficients $\kappa_\|$ and $\kappa_\perp$ by a generalized Einstein relation, ensuring the asymptotic approach to kinetic equilibrium independently from the discretization scheme employed in the numerical implementation of Eq.~(\ref{eq:Langevin})}.

\bibliographystyle{iopart-num}
\bibliography{paper}

\end{document}